\documentclass[%
aps,
pre,
 amsmath,amssymb,
 superscriptaddress,
 preprint,%
longbibliography
]{revtex4-1}

\usepackage{graphicx}
\usepackage{dcolumn}
\usepackage{bm}
\usepackage[mathlines, modulo]{lineno}
\usepackage{ulem}
\usepackage[utf8]{inputenc}
\usepackage[T1]{fontenc}
\usepackage{mathptmx}
\usepackage{textcomp} 
\usepackage{nicefrac}
\usepackage{xcolor}
\usepackage{gensymb}
\usepackage{multibib}
\usepackage{hyperref}

\newcommand{\ve}[1]{\boldsymbol{#1}}

\DeclareMathAlphabet{\pazocal}{OMS}{zplm}{m}{n}

\newcommand{\wse}{\ifmmode \mathrm{WSe\textsubscript{2}}\else WSe\textsubscript{2}\fi}
\newcommand{\ws}{\ifmmode \mathrm{WS\textsubscript{2}}\else WS\textsubscript{2}\fi}
\newcommand{\mos}{\ifmmode \mathrm{MoS\textsubscript{2}}\else MoS\textsubscript{2}\fi}
\newcommand{\mose}{\ifmmode \mathrm{MoSe\textsubscript{2}}\else MoSe\textsubscript{2}\fi}

\begin{document}

\title{Probing excitons with time-resolved momentum microscopy}

\author{Marcel Reutzel} \email{marcel.reutzel@phys.uni-goettingen.de}%
\address{I. Physikalisches Institut, Georg-August-Universit\"at G\"ottingen, Friedrich-Hund-Platz 1, 37077 G\"ottingen, Germany}

\author{G.~S.~Matthijs~Jansen} 
\email{gsmjansen@uni-goettingen.de}
\address{I. Physikalisches Institut, Georg-August-Universit\"at G\"ottingen, Friedrich-Hund-Platz 1, 37077 G\"ottingen, Germany}

\author{Stefan Mathias} 
\email{smathias@uni-goettingen.de}%
\address{I. Physikalisches Institut, Georg-August-Universit\"at G\"ottingen, Friedrich-Hund-Platz 1, 37077 G\"ottingen, Germany}
\address{International Center for Advanced Studies of Energy Conversion (ICASEC), University of Göttingen, Göttingen, Germany}

\begin{abstract}

Excitons - two-particle correlated electron-hole pairs - are the dominant low-energy optical excitation in the broad class of semiconductor materials, which range from classical silicon to perovskites, and from two-dimensional to organic materials. Recently, the study of excitons has been brought on a new level of detail by the application of photoemission momentum microscopy -  a technique that has dramatically extended the experimental capabilities of time- and angle-resolved photoemission spectroscopy (trARPES). Here, we review how the energy- and momentum-resolved photoelectron detection scheme enables direct access to the energy landscape of bright and dark excitons, and, more generally, to the momentum-coordinate of the exciton that is fundamental to its wavefunction. Focusing on two-dimensional materials and organic semiconductors as two tuneable platforms for exciton physics, we first discuss the typical photoemission fingerprint of excitons in momentum microscopy and highlight that is is possible to obtain information not only on the electron- but also hole-component of the former exciton. Second, we focus on the recent application of photoemission orbital tomography to such excitons, and discuss how this provides a unique access to the real-space properties of the exciton wavefunction. Throughout the review, we detail how studies performed on two-dimensional transition metal dichalcogenides and organic semiconductors lead to very similar conclusions, and, in this manner, highlight the strength of time-resolved momentum microscopy for the study of optical excitations in semiconductors.

\end{abstract}

\maketitle


\section{Introduction}

Angle-resolved photoemission spectroscopy (ARPES) provides access to the binding energy $E$ and the momentum $\ve{k}$ of photoelectrons in a condensed matter material. The photoelectron distribution therefore contains direct information on the equilibrium single-particle band structure, and also on many-body correlation effects via the electronic self energy~\cite{hufner_photolectron_2003, Damascelli03rmp}. In order to access the non-equilibrium properties of a material after an external perturbation, ARPES experiments can be performed in a pump-probe scheme. In time-resolved ARPES (trARPES)~\cite{Bovensiepen12book, Sobota21rmp, Boschini2023arxiv}, a pump laser pulse photoexcites the sample before a time-delayed probe laser pulse is used to detect the energy- and in-plane momentum-resolved photoemission spectrum. Very generally, the optical excitation creates a non-equilibrium distribution of electrons and holes that respond to the screening cloud of surrounding quasiparticles and dissipate their excess energy in secondary scattering processes. For materials with a high charge carrier density, the Coulomb attraction between the single-particle electrons and holes is efficiently screened, and the non-equilibrium distribution of single-particle electrons and holes thermalizes by carrier-carrier and carrier-phonon scattering events. In this manner, trARPES (aka time-resolved two-photon photoemission spectroscopy (tr2PPE)) has been successfully applied to a huge class of materials, ranging from pristine and adsorbate covered surfaces~\cite{Echenique04ssr, Bauer15pss, Petek23pss, Zhu04ssr} to quantum materials such as graphene~\cite{Johannsen13prl, Gierz13natmat,Na19sci, Duvel22nanolett}, topological insulators~\cite{Sobota12prl, Niesner14prb, Reimann14prb,mahmood_selective_2016} and the melting of charge-ordered states~\cite{Schmitt08sci, Rohwer:2011fy}.

For semiconductors that are characterized by a small charge carrier density and a weak dielectric screening, the Coulomb interaction between the optically excited single-particle electrons and holes is not efficiently screened and correlated electron-hole two-particle states with significant binding energy E$_{\rm bin}$ - so-called excitons - are formed~\cite{haug2009quantum, kira2011semiconductor, Huber01nat, Wang18rmp}. Excitons can not be described in the single-particle band structure, and they manifest as additional peaks in absorption spectra at energies of one exciton binding energy below the single-particle conduction band. Naturally, being the lowest energy excitation, excitons then dominate the optical response of the wide class of semiconductors that ranges from classical silicon to perovskites, two-dimensional transition metal dichalcogenides (TMDs) and organic semiconductors. In consequence, it is crucial to understand the excitonic response of semiconductors to an optical excitation on a fundamental level.

Given that excitons represent electronic excitations that are usually in the visible frequency range, their excitation energies and relaxation dynamics are often studied with methods such as (time-resolved) absorption or fluorescence spectroscopy. As a photon-in-photon-out technique, besides notable exceptions~\cite{Kaindl03nat, Poellmann15natmat, Merkl19natmat}, all-optical techniques are only sensitive to transitions that couple to the light field. Fluorescence, for example, can only be used to observe excitons that can decay under to the emission of a photon in a radiative process and thus have a measurable oscillator strength~\cite{Kira99pqe, Koch06natmat, Nayak17acsnano}. Hence, so-called dark excitons, where the electron- and the hole-component have a finite center-of-mass momentum or are of different spin, can only be accessed if an additional scattering event with a phonon or a spin-flip occurs~\cite{Malic18prm, Wu18prb, Lindlau18natcom,Brem20nanolettersPhononAssisted}. Consequently, these optical spectroscopies provide no information on the momentum of the exciton, which, however, is key to the full description of the exciton wavefunction~\cite{Rustagi18prb}. In contrast, as a photon-in-electron-out technique, trARPES provides direct access to the energy- and the momentum-coordinate of the photoemitted single-particle electrons that originate from the break-up of two-particle excitons. Therefore, momentum-indirect and spin-forbidden (optically dark) excitons can be accessed in a trARPES experiment.

While time-resolved photoemission spectroscopy has already been applied to various types of excitons in different material classes~\cite{Weinelt04prl, Varene12prl, Deinert14prl, Chan11sci, Cui14natphys, Tanimura19prb, Tanimura23prb, Zhu15jesrp, Mori23nat, gierster2023arxiv}, recent developments on the side of the femtosecond laser light sources, the photoelectron analyzer, and also the material aspect have established trARPES as one of the most powerful tools to study exciton dynamics~\cite{Madeo20sci, Wallauer21nanolett, Man21sciadv, Dong20naturalsciences, Schmitt22nat, Bange232DMaterials, kunin23prl, bange24SciAdv, Schmitt23arXiv, Karni22nat, Bennecke23arxiv, Neef23nat, beaulieu2023berry, Karni23advmat, Liu23cs}. First, as mechanical exfoliation of two-dimensional materials nowadays allows for atomic-scale control of artificial heterostructures~\cite{Geim11nat}, several materials have become available that show strong excitonic response and a large exciton  binding energy in the range of a few 100~meV~\cite{Chernikov14prl, He14prl}. This has made excitonic systems more accessible to trARPES, as the large binding energy enables to discriminate photoelectrons originating from excitons or from the conduction band. Second, the rapid development of photoelectron spectroscopy detectors has culminated in the development of the photoelectron momentum microscope, which enables efficient detection of full 2D momentum-, kinetic energy-, and even spin-resolved photoelectron spectra~\cite{kromker_development_2008, Tusche15ultra, medjanik_direct_2017, kutnyakhov_time-_2020}. Third, the development of momentum microscopes was complemented by the development of high-repetition rate femtosecond extreme ultraviolet light sources~\cite{Heyl12jpb, carstens_high-harmonic_2016, chiang_boosting_2015, Keunecke20timeresolved, moller_ultrafast_2021, puppin_time-_2019}. Here, the high repetition rate enables an efficient combination with momentum microscopes, where only about 1 electron/pulse can be detected. At the same time, time-resolved momentum microscopy becomes possible with simultaneous access to electrons with large in-plane momenta >1~\AA$^{-1}$~\cite{mathias_angle-resolved_2007}, as is necessary to probe the full Brillouin zone of typical two-dimensional materials and the orbital fingerprints of organic semiconductors. This is highly beneficial for exciton spectroscopy, as the multidimensional resolution directly allows to distinguish between various optically bright and dark excitonic states, and even allows access to the spatial exciton wavefunction in two-dimensional and organic semiconductors~\cite{Man21sciadv, Schmitt22nat, Dong20naturalsciences, Baumgartner22natcom, Neef23nat, Wallauer20sci, Adamkiewicz23jpcc, Bennecke23arxiv} within the framework of orbital tomography~\cite{Puschnig09sci}. Given this rapid progress in recent years, here, we present a review of the recent efforts in this research field to characterize excitons in two-dimensional and organic semiconductors. Importantly, we do this from the perspective of the momentum-resolved measurement scheme and discuss how this information can be exploited to reconstruct information on the real-space wavefunction of the optical excitation, i.e., the exciton.

The review article is structured as follows: In section \ref{sec:theory}, we review the basic theory of excitons, and answer the question how it is, in theory, possible to probe an exciton by measuring single-particle photoelectrons. We also present a concise overview of the recently developed capabilities of photoemission momentum microscopy. In section \ref{sec:excitons-in-pes}, we will first focus on the impact that momentum microscopy has had on the study of excitons in two-dimensional TMDs and organic semiconductors. Thereby, we will discuss how the energy- and momentum-resolved photoelectron detection can be used to quantify the energy landscape of excitons and how distinct signatures of the exciton's hole-component are found in the photoemission signal. In section \ref{sec:progressOS}, we outline how the framework of orbital tomography has been used to reconstruct the real-space wavefunction of excitons in TMDs and organic semiconductors.


\section{Methods: probing excitons with momentum microscopy}
\label{sec:theory}

\begin{figure}[b]  
    \centering
    \includegraphics[width=.99\linewidth]{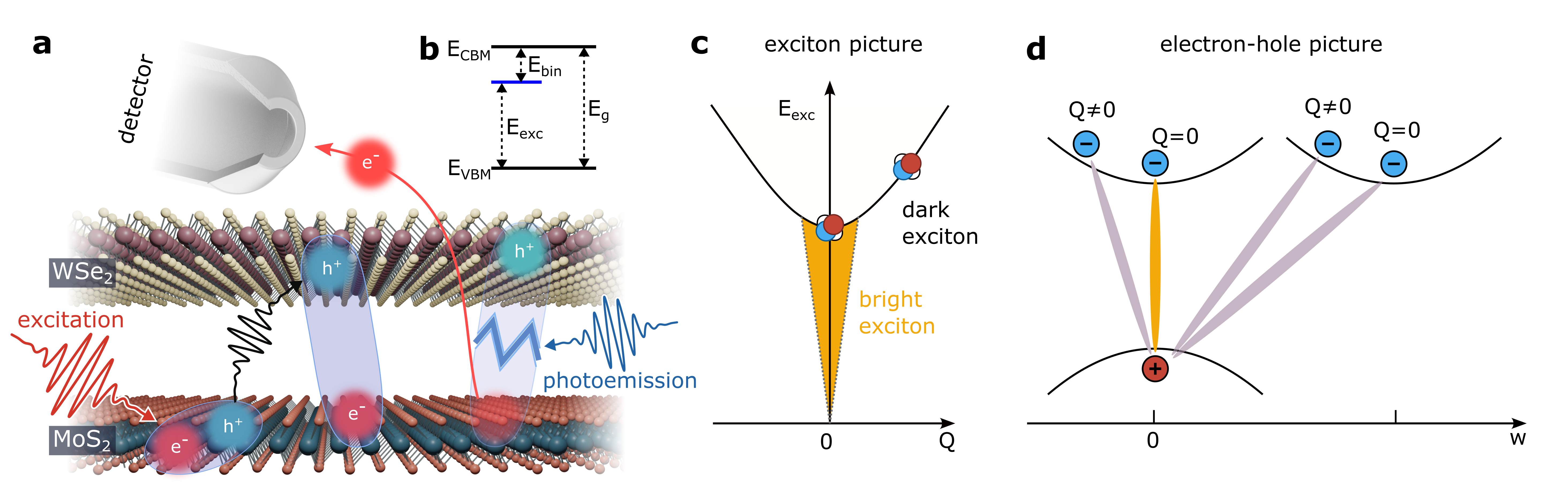}
    \caption{Description of excitons and their detection in a time- and angle-resolved photoemission spectroscopy experiment.
    \textbf{a} Schematic illustration of the exciton formation, thermalization and detection process. A pump laser pulse (red) is used to excite optically bright excitons that reside in a single TMD layer. Interlayer charge-transfer can lead to the formation of interlayer excitons where the exciton's electron and hole reside in different layers. In the photoemission process, the Coulomb correlation between the electron and the hole is broken, the single-particle electron is detected with the photoelectron analyzer and the single-particle hole remains in the sample.
    \textbf{b} Schematic energy level diagram indicating the single-particle valence band maximum (or HOMO) and conduction band minimum (or LUMO) at binding energies $E_{\rm VBM}$ and $E_{\rm CBM}$ that are separated by the single-particle band gap energy E$_g$, respectively. In this picture, the binding energy $E_{\rm bin}$ of a two-particle excitons can be defined by comparing the exciton energy E$_{\rm exc}$ with the single-particle band gap E$_g$.
    \textbf{c} In the exciton picture (shown for $\ve{w}=0$), Coulomb correlated electron-hole pairs are described based on equation~(\ref{eq:excitonenergy}) and have a parabolic dispersion with regard to their center-of-mass momentum $Q$. Within the light cone (orange area, vanishing $Q$), excitons are labeled to be optically bright because they can be excited by light and can decay in a radiative process. In contrast, excitons with a finite $Q$ are termed optically dark.
    \textbf{d} A related pictorial description of excitons can be drawn in the electron-hole picture. Here, in addition, momentum-indirect excitons are sketched where the electron- and the hole-component are separated by momentum $\ve{w}$ and reside in different valleys of the Brillouin zone.
    Panel \textbf{a} is reproduced from ref.~\cite{bange24SciAdv} under Creative Commons Attribution License 4.0 (CC BY). Panels \textbf{c,d} are adopted from ref.~\cite{Werner23ma}.
    }
    \label{fig:break-exciton}
\end{figure}

The first question that comes to mind when discussing photoemission spectroscopy of excitons is: how does an exciton show-up in an energy- and momentum-resolved photoemission spectrum? This question follows naturally from the basic description of an exciton: an exciton is an excited state of the many-body electron wavefunction of a semiconductor that can be conveniently described as a quasiparticle consisting of a Coulomb-correlated electron-hole pair. Thus, the theoretical description of an exciton requires a description of the many-body electron wavefunction, while in trARPES, single-particle photoelectrons are emitted and detected. In a simplistic picture, the photoemission process can be described as follows (Fig.~\ref{fig:break-exciton}a): After a light pulse with photon energy $\hbar\omega_{\rm pump}$ (red) excites the semiconductor and creates an exciton, the photoemission laser pulse $\hbar\omega_{\rm probe}$ (blue) excites the already-excited electron to a free-electron-like state, and thereby breaks the Coulomb-correlated electron-hole pair. The resulting single-particle photoelectron then travels to the surface, leaves the sample, and is detected with the photoelectron analyzer, while the single-particle hole remains in the sample. This simplified scenario provides an intuitive representation of the photoemission process, however, also a series of major questions arise: Very generally, in how far does the trARPES experiment provide access to information on the exciton's electron and hole components and also the correlation between both? More specifically, at which single-particle photoelectron energy is the former exciton's electron detected? Which information is contained in the momentum-resolved photoemission signal? Can the framework of orbital tomography be applied to reconstruct the real-space distribution of the exciton wavefunction?

In the following, we will first briefly describe the current state-of-the art of time-resolved momentum microscopy experiments. Next, we will review theoretical proposals on how to detect photoelectron signal from excitons.


\subsection{Femtosecond momentum microscopy - a new variant of trARPES}

\begin{figure}[b]  
    \centering
    \includegraphics[width=.95\linewidth]{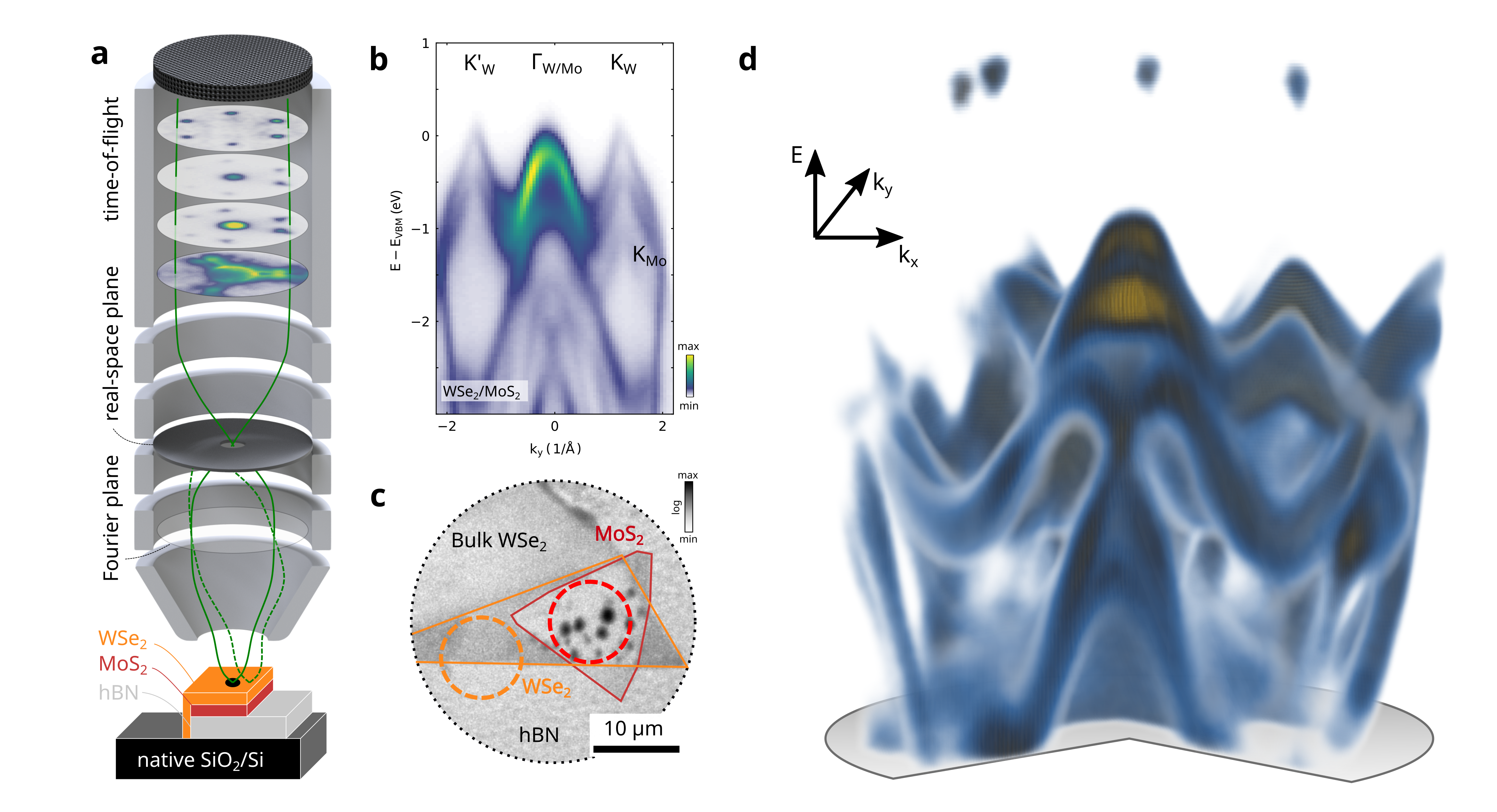}
    \caption{Schematic illustration of the momentum microscopy setup and the accessible multi-dimensional photoemission data.
    \textbf{a} Momentum microscopes are a novel type of photoelectron analyzers that are assembled by a microscope type electronic lens system, a time-of-flight drift tube or a hemisphere-based energy filter, and a position sensitive photoelectron detector. By projecting either the Fourier or the real-space plane onto the detector, spectrally-resolved momentum maps (\textbf{b}) or real-space maps of the sample (\textbf{c}) can be collected. By inserting an aperture into the real-space plane, a region-of-interest with a diameter of approximately 10~$\mu$m can be selected on the sample.
    \textbf{d} If the momentum microscope is equipped with a time-of-flight detector, it is possible to simultaneously collect three-dimensional data sets that contain information on the kinetic energy $E$ and two in-plane momenta (k$_x$, k$_y$) of the photoelectrons. The 3-dimensional data representation shows the occupied bands of homobilayer MoS$_2$ and excitonic photoemission signal at a pump-probe delay of 140~fs. 
    Panels \textbf{a}, \textbf{b}, and \textbf{c} are reproduced from ref.~\cite{Schmitt22nat} (Copyright by Springer Nature). The data in panel \textbf{d} was taken in the Göttingen photoemission laboratory.  
    }
    \label{fig:experimentalsetup}
\end{figure}

Time- and angle-resolved photoemission spectroscopy (trARPES) is a powerful technique to map the electronic band structure, identify many-body interactions and quantify ultrafast energy dissipation processes under equilibrium and non-equilibrium conditions~\cite{Damascelli03rmp, Bovensiepen12book, Sobota21rmp, Boschini2023arxiv, gudde07sci, mahmood_selective_2016, Rohwer:2011fy, Schmitt08sci, Reutzel20natcom}. Hence, with the fabrication of two-dimensional monolayers and their heterostructures, the idea to apply trARPES to study their non-equilibrium response has been straightforward. At the same time, it has been highly desirable to extend the framework of orbital tomography~\cite{Puschnig09sci} into the non-equilibrium regime in order to reconstruct orbitals of initially unoccupied states~\cite{Wallauer20sci} or even excitons~\cite{Kern23prb, Bennecke23arxiv, Neef23nat,Karni22nat, Schmitt22nat, Dong20naturalsciences, Man21sciadv}. For the case of van-der-Waals semiconductors, early experiments could address bulk and wafer-scale materials~\cite{Bertoni16prl, Wallauer16apa, Hein16prb, Wallauer20prb, Dong20naturalsciences, Grubisic15nanolett, Liu19prl, Liu20prb, Lin22prb, Lee21nanolett, Aeschlimann20sciadv}, however, major experimental challenges had to be overcome before high-quality exfoliated van-der-Waals heterostructures could be studied~\cite{Madeo20sci, Man21sciadv, Wallauer21nanolett, kunin23prl, Schmitt22nat, Bange232DMaterials, bange24SciAdv, Schmitt23arXiv, Dong23natcom, Potocnik23nanolett}. The first experimental challenge lies in the small diameter of the exfoliated sample systems and the limited spatial resolution of common trARPES experiments. Second, extreme ultraviolet photon pulses at several hundreds of kHz repetition rate needed to be generated before high-quality momentum microscopy experiments could be reported.

As an integrating technique, the spatial resolution of established trARPES experiments equipped with hemispherical energy analyzers is determined by the spot sizes of the probe laser pulse on the surface. While typical laser beam diameters are in the range of 100~$\mu$m and more, highest quality exfoliated two-dimensional materials have diameters of roughly 10~$\mu$m. Hence, it is not possible to study such samples with established laser-based table-top experiments. In static ARPES experiments performed at synchrotron light sources, this challenge has been overcome by using Fresnel zone plates or capillary devices to reduce the beam diameter on the sample to the sub-1-$\mu$m regime~\cite{Cattelan18nanomaterials, Rotenberg14jsr}. By scanning the laser-spot across the sample, the occupied electronic band structure and many-body interaction effects of mono-, homo- and heterolayers of two-dimensional materials have been characterized~\cite{Wilson17sciadv, Ulstrup20sciadv, Lisi21natphys}. It has even been possible to perform ARPES experiments on electrostatically gated samples~\cite{Nguyen19nat, Hofmann21avsquantumscience}. While such experiments are fully established and well-known under the acronyms $\mu$ARPES and nanoARPES, until the recent addition of momentum microscopes in ARPES experiments, it has not been possible to extend $\mu$ARPES and nanoARPES towards femtosecond time-resolved experimental schemes using extreme ultraviolet light pulses.

We want to emphasize that the usage of extreme ultraviolet pulses is indispensable for the study of exciton dynamics in two-dimensional and organic materials. This is because of restrictions in the photoemission process~\cite{hufner_photolectron_2003, mathias_angle-resolved_2007, Li20njp} that require photon energies larger than 20~eV in order to probe large in-plane momenta. In materials such as graphene and TMDs, for example, the K valleys are found at $\approx1.7$\AA$^{-1}$ and $\approx1.2$\AA$^{-1}$, respectively, and are not accessible with usual femtosecond laser sources. Likewise, without extreme ultraviolet light pulses, it would not be possible to map the dynamics of the momentum fingerprint of a molecular orbital, as necessary for the evaluation of the trARPES data in the framework of photoemission orbital tomography~\cite{Puschnig09sci, jansen_efficient_2020, Wallauer20sci, Bennecke23arxiv, Kern23prb}. With the development of high-repetition rate high-harmonic generation beamlines~\cite{Heyl12jpb, carstens_high-harmonic_2016, chiang_boosting_2015,Keunecke20timeresolved, moller_ultrafast_2021} and the newest generation free-electron lasers~\cite{decking_mhz-repetition-rate_2020, faatz_simultaneous_2016}, it is possible to generate bright EUV light pulses with pulse durations in the attosecond to few-femtosecond range. However, the added complexity of (sub-)micron-scale focusing at EUV wavelengths has so far hindered focusing these pulses to achieve time-resolved EUV $\mu$ARPES and nanoARPES. 

Nevertheless, the last few years have shown that a complementary photoelectron detection scheme is extraordinarily well suited for trARPES of exfoliated samples: Momentum microscopes~\cite{kromker_development_2008, medjanik_direct_2017} represent a family of photoelectron spectroscopy devices that are based on the electrostatic lens system of a (photoemission) electron microscope (PEEM). Crucially, this lens system can be controlled such that not only the real-space image plane can be projected onto the detector, but as well the Fourier plane (i.e., momentum space) is accessible (Fig.~\ref{fig:experimentalsetup}a). Combined with spectral selection, e.g. via time-of-flight, and position-sensitive photoelectron detector, momentum microscopes then provide three-dimensional (3D) access to the photoemission intensity as a function of the photoelectron energy $E$ and two in-plane momenta $k_x$ and $k_y$~\cite{medjanik_direct_2017, Keunecke20timeresolved, kutnyakhov_time-_2020, Maklar20rsi, Keunecke20prb, Li22prb} (Fig.~\ref{fig:experimentalsetup}d) or the photoelectron energy and two real-space coordinates $r_x$ and $r_y$~\cite{Barrett12rsi, Fujikawa09prb, Schmitt23arXiv}. In order to access exfoliated samples with a diameter of 10~$\mu$m, it is then straightforward to insert an aperture into the real-space plane of the microscope and to measure only the photoelectrons coming from the selected region of interest. Note that, depending on the available apertures and lens settings, this approach also allows to set regions of interest with a diameter less than 10~$\mu$m. Naturally, this then increases the measurement time.

The strategy to perform femtosecond momentum microscopy experiments on exfoliated van-der-Waals materials is then as follows: First, in the real-space mode of the microscope, the sample structure can be mapped based on the photoemission intensity contrast. As an example, the orange and the dark red polygons in Fig.~\ref{fig:experimentalsetup}c indicate monolayers of WSe$_2$ and MoS$_2$, respectively. Subsequently, three-dimensional ARPES spectra can be collected by inserting the real-space aperture in the respective region of interest (orange and red circles in Fig.~\ref{fig:experimentalsetup}c) and then projecting the Fourier plane onto the photoelectron detector (Fig.~\ref{fig:experimentalsetup}b, d). In combination with a pump-probe scheme,  multi-dimensional data sets of the temporal evolution of the non-equilibrium dynamics of quasiparticles can be measured.

\subsection{The exciton wavefunction}
\label{subsec:exciton_wavefunction}

Before it is possible set up a theory of photoemission from excitons, it is first necessary to place the excitonic wavefunction on a theoretical basis. An exciton is essentially caused by a renormalization of the many-body electron wavefunction after an optical excitation, which requires by definition a step beyond ground-state theory. Whereas ground-state electronic structure calculations such as density functional theory (DFT) or the many-body interaction-corrected $GW$ framework provide a good description of the valence and conduction orbitals $\phi_v(r)$ and $\chi_c(r)$ and their corresponding energies of the unexcited system, these theories ultimately describe single-quasiparticle orbitals and do not directly provide a good indication of either the exciton energy spectrum or the associated exciton wavefunctions. 
Rather, the wavefunction of an exciton must generally be expressed as a linear combination of electron-hole pairs $\phi_v(r_h)\chi_c(r_e)$. Instead of a single (electron) coordinate, this expression depends on the real-space coordinates $r_e$ and $r_h$ of the electron and hole wavefunctions, respectively.
In the Tamm-Dancoff approximation \cite{Rohlfing00}, the wavefunction of the $m$\textsuperscript{th} exciton can be written as
\begin{equation}
    \psi_m(\ve{r}_h, \ve{r}_e) = \sum_{v,c} A^{(m)}_{vc} \phi_v^*(\ve{r}_h) \chi_c(\ve{r}_e).
    \label{eq:exciton_state}
\end{equation}
Here, the expansion coefficients $A^{(m)}_{vc}$ describe the relative contribution of each electron-hole-pair to the exciton wavefunction. Equivalently, $|A^{(m)}_{vc}|^2$ indicates the probability to simultaneously find an electron in the conduction orbital $\chi_c$ and a hole in the valence orbital $\phi_v$. The exciton energies $E_{\rm exc}^{(m)}$ and the corresponding coefficients $A^{(m)}_{vc}$ can be calculated by several methods~\cite{Hagel21prr, katsch2018theory}. Notably, solving the Bethe-Salpeter equation enables a fully-interacting calculation of the exciton spectrum~\cite{Rohlfing00, Blase2020, quintela_theoretical_2022}, and time-dependent density functional theory (TD-DFT) using the Casida formalism provides a powerful and much used alternative~\cite{casida_time-dependent_1995, andrade_real-space_2015, tancogne-dejean_octopus_2020}. 

In principle, the full exciton wavefunction (equation~\eqref{eq:exciton_state}) includes excitation and de-excitation contributions of all valence and conduction orbitals (bands) in the material, which can challenge the intuitive understanding of the exciton wavefunction. However, often the situation is not so complex: many exciton states in semiconducting TMDs, in particular those of low energy, are described by a hole in a single valence and an electron in a single conduction band (Fig.~\ref{fig:break-exciton}b,c). Similarly, in organic semiconductors, the lowest excitonic states are usually composed of an electron in the lowest unoccupied molecular orbital (LUMO) and a hole in the highest occupied molecular orbital (HOMO). 


\subsection{Excitons in two-dimensional semiconductors}
\label{subsec:excitons_in_2D_theory}

To understand the benefits of trARPES for the study of excitons, it is first useful to consider the prominent case of excitons in the family of semiconducting TMDs, which include \ws{}, \wse{}, \mos{} and \mose{}~\cite{Wang18rmp}. These materials already show a significant excitonic response in the bulk phase, but they are especially interesting in their monolayer form: here, a direct band gap forms at the K valleys~\cite{Mak10prl, Splendiani10nanolett}, which, among others, host the bright A1s-exciton~\cite{Chernikov14prl, He14prl}.

For the A1s-exciton, the electron orbitals are given by the quasi-parabolic conduction band at the K valley, while the hole orbitals are given by the negative quasi-parabolic valence band at the K valley. The interaction between electron and hole is given by the Coulomb interaction. Consequentially, the exciton can be described by a two-dimensional analogue of the hydrogen atom. For such an exciton, the exciton energy  $E_{\rm exc}$ can be expressed as 
\begin{equation}
    E_{\rm exc}\left(\ve{Q}\right)=E_g-E_{\rm bin}+\frac{\ve{Q}^2}{2M}. 
    \label{eq:excitonenergy}
\end{equation}
where $E_g$ is the single-particle band gap (Fig.~\ref{fig:break-exciton}b) and the exciton mass $M$ can be calculated from the effective mass of the electron and the hole in the conduction and valence bands, respectively. As in the case of the hydrogen atom, the exciton binding energy E$_{\rm bin}$ depends on the quantum number $n$, and, for the lowest energy exciton $n=1$, can be on the order of 0.5~eV~\cite{Chernikov14prl, mueller18}. The last term of equation~\eqref{eq:excitonenergy} carries information on the kinetic energy of the exciton, which is parameterized by the kinetic center-of-mass (COM) momentum $\ve{Q}$.

If a direct semiconductor with a single valley is optically excited on resonance with the exciton energy $E_{\rm exc}$, the momenta of the exciton's electron and hole are $\ve{k}_e$ and $-\ve{k}_h$, respectively, with $|\ve{k}_e|$ = $|-\ve{k}_h|$ such that $\ve{Q}=\ve{k}_e+\ve{k}_h=0$ (Fig.~\ref{fig:break-exciton}c). In the case that $\ve{Q}\neq0$, the electron-hole pairs are shifted out of the light-cone, and, because momentum needs to be conserved in a radiative decay process, they are not anymore straightforwardly accessible in all-optical experiments. These excitons are therefore termed to be optically dark. In addition to such intravalley excitons, for two-dimensional semiconductors, the thermalization of the optical excitation is strongly determined by intervalley scattering and the formation of momentum-indirect excitons where the electron- and hole-components are in different valleys of the Brillouin zone~\cite{Malic18prm, Zhang15prl,Rivera16sci, Rivera18natnano}. As illustrated in Fig.~\ref{fig:break-exciton}d, this includes the formation of excitons where the electron- and the hole-component reside in different valleys of the Brillouin zone, separated by momentum $\ve{w}$. For those intervalley excitons, the COM is then described by $\ve{Q}=\ve{k}_e+\ve{k}_h-\ve{w}$, and these excitons are also not straightforwardly accessible in an all-optical experiment. 


\subsection{Breaking the exciton in the photoemission process: Energy- and momentum-conservation}
\label{subsec:E_k_conservation_theory}
In the photoemission process, the probe laser pulse breaks the exciton into its single-particle components. The former exciton's electron is detected by the photoelectron analyzer and the former exciton's hole remains in the sample (Fig.~\ref{fig:break-exciton}a). In this process, the energy and the momentum of the total system need to be conserved. For excitons with a center-of-mass momentum $\ve{Q}=0$, Weinelt \textit{et al.}~\cite{Weinelt04prl} formulated it as follows: Before and after the breakage of the exciton, the total energy of the system is given by $E_{\rm tot}=E_0+E_{\rm exc}+\hbar\omega_{\rm probe}$ and $E_{\rm tot}=E_0-E_{\rm hole}+E_{\rm elec}$, respectively. Here, $E_0$ is the ground state energy before the optical excitation and $E_{\rm hole}$ ($E_{\rm elec}$) are energies of the single-particle holes (electrons) after the breakup of the exciton. 
The energy of the single-particle photoelectrons can then be expressed as
\begin{equation}
    E_{\rm elec}=E_{\rm hole}+E_{\rm exc}+\hbar\omega_{\rm probe}.
    \label{eq:enmomconservation}
\end{equation}
Equation~\eqref{eq:enmomconservation} directly shows that the energy- and the momentum-coordinate of the photoemitted electrons $E_{\rm elec}$ depends on the energy and the momenta of the single-particle hole that remains in the sample $E_{\rm hole}$: The photoelectrons are detected one exciton energy $E_{\rm exc}$ above the valence bands whereas the single-particle hole remains in the sample. 
For the case that the exciton is composed of a single pair of valence and conduction bands, the result of equation~\eqref{eq:enmomconservation} can be intuitively expressed in terms of the exciton binding energy $E_{\rm bin}$ (Fig.~\ref{fig:break-exciton}b): At the cost of the exciton binding energy $E_{\rm bin}$, single-particle photoelectrons originating from excitons are found below the conduction band minimum (i.e., $E_{\rm VBM}+E_{\rm exc}=E_{\rm CBM}-E_{\rm bin}$). 

\begin{figure}[bt]  
    \centering
    \includegraphics[width=.8\linewidth]{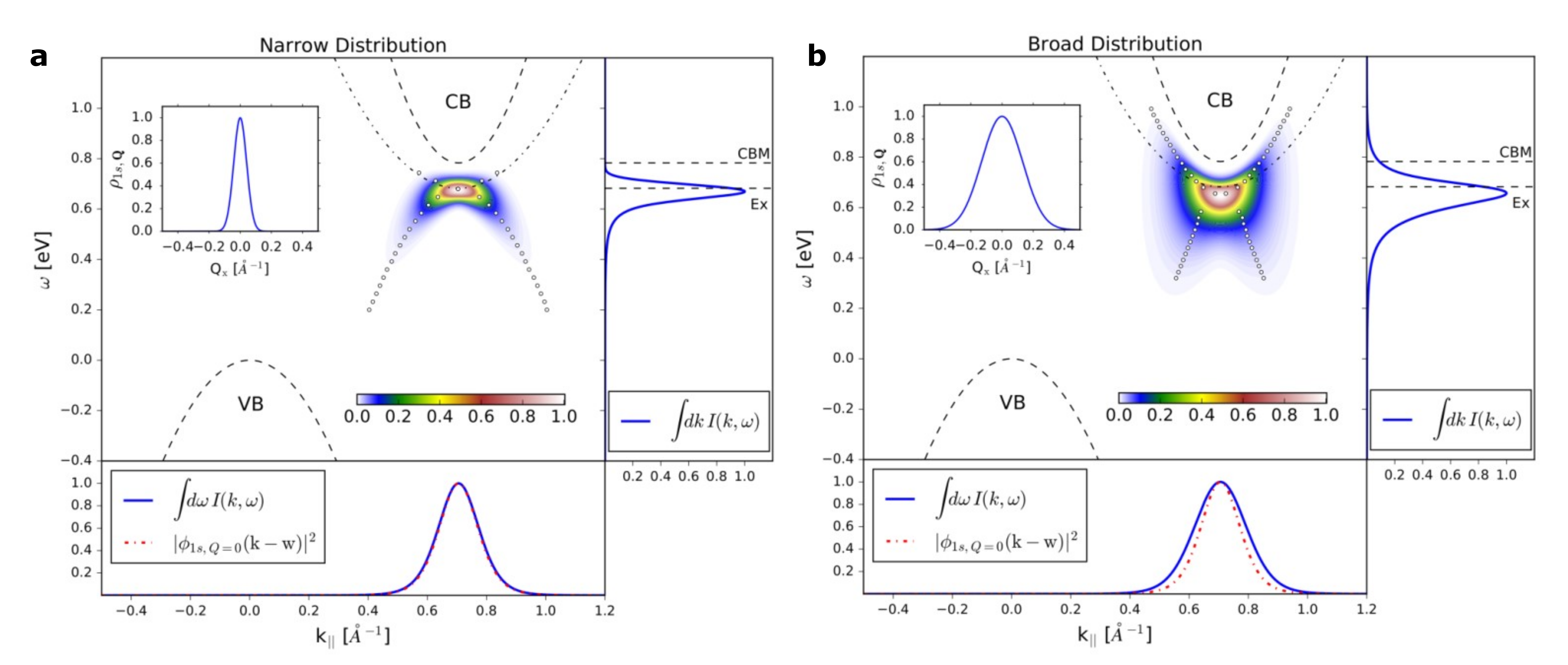}
    \caption{The photoemission signature from excitons is affected by the joint electron-hole nature of the compound quasiparticle. 
    \textbf{a} For cold, $Q=0$ excitons in TMDs, it was found that the dispersion of the hole state is directly imprinted on the trARPES data. 
    \textbf{b} For hot excitons including a wider range of momenta $Q$ (see insets), the trARPES signature is broadened and the hole-dispersion is no longer apparent. 
    Figure reproduced from ref.~\cite{Rustagi18prb}. Copyright 2018 by the American Physical Society}
    \label{fig:photoemission-theory}
\end{figure}

This connection between the electron- and the hole-component of the exciton in the photoemission process is now formulated by multiple groups on a different level of theory~\cite{Ohnishi18, Rustagi18prb, Christiansen19prb, Steinhoff2017natcom, Meneghini23ACSPhotonics, Kern23prb, Perfetto16prb}. Figure~\ref{fig:photoemission-theory} exemplarily shows the analysis of Rustagi and Kemper~\cite{Rustagi18prb} for the case of an indirect semiconductor where the exciton's electron and hole component reside in different valleys ($\ve{w} \neq 0$). In the limit of a low exciton density where the excitons have no center-of-mass momentum (Fig.~\ref{fig:photoemission-theory}a), the photoemission from the exciton is found one exciton energy above the valence band (respectively, one exciton binding energy below the conduction band). Importantly, the energy-momentum dispersion shows the hole-like dispersion of the valence band, as expected from energy- and momentum-conservation following equation~(\ref{eq:enmomconservation}). Notably, the situation is different if the exciton occupation is hot and excitons with a finite center-of-mass momentum contribute (Fig.~\ref{fig:photoemission-theory}b). In this scenario, the pure valence band like dispersion of the excitonic photoemission signal is not anymore observed. However, the photoemission signal is still found one exciton binding energy E$_{\rm bin}$ below the conduction band, facilitating the discrimination of photoelectrons being emitted from excitons or the conduction band.


\subsection{Generalization to multi-orbital excitons}
\label{subsec:multiorbital_theory}

So far, excitons have been considered that are derived from a single pair of valence and conduction bands, such as sketched in Fig.~\ref{fig:break-exciton}d. In this case, the connection between the exciton binding energy $E_{\rm bin}$, the exciton energy $E_{\rm exc}$ and the single particle valence and conduction band energies is straightforward (Fig.~\ref{fig:break-exciton}b). However, as described by equation~\eqref{eq:exciton_state}, an exciton can be composed of multiple orbital pairs, where the single-particle electrons and holes reside in different single-particle conduction and valence bands, respectively. 

Equation~\eqref{eq:exciton_state} can be understood as the coherent sum of electron-hole pairs, where the exciton's electron and hole component reside in different conduction and valence bands. It is known that such multi-orbital excitons strongly contribute to the optical properties of organic semiconductors~\cite{krylov_orbitals_2020, martin_natural_2003, hammon_pump-probe_2021, Kern23prb, Bennecke23arxiv}, but they can also contribute to the optical response of multilayer TMD systems, where interlayer hybridization effects lead to the formation of valence and conduction bands with bonding and anti-bonding character~\cite{Meneghini23ACSPhotonics,Cheiwchanchamnangij12prb, Splendiani10nanolett}. In such a complex scenario, it is not directly clear at which energy and momentum the photoelectrons originating from the excitons are expected. While time-dependent density functional theory methods provide a powerful and generally applicable approach to achieve this information~\cite{degiovanni_first-principles_2017}, a more direct theoretical description would be beneficial, and was recently developed in two publications by Kern \textit{et al.}~\cite{Kern23prb} and Meneghini \textit{et al.}~\cite{Meneghini23ACSPhotonics}.

\begin{figure}[bt]  
    \centering
    \includegraphics[width=.8\linewidth]{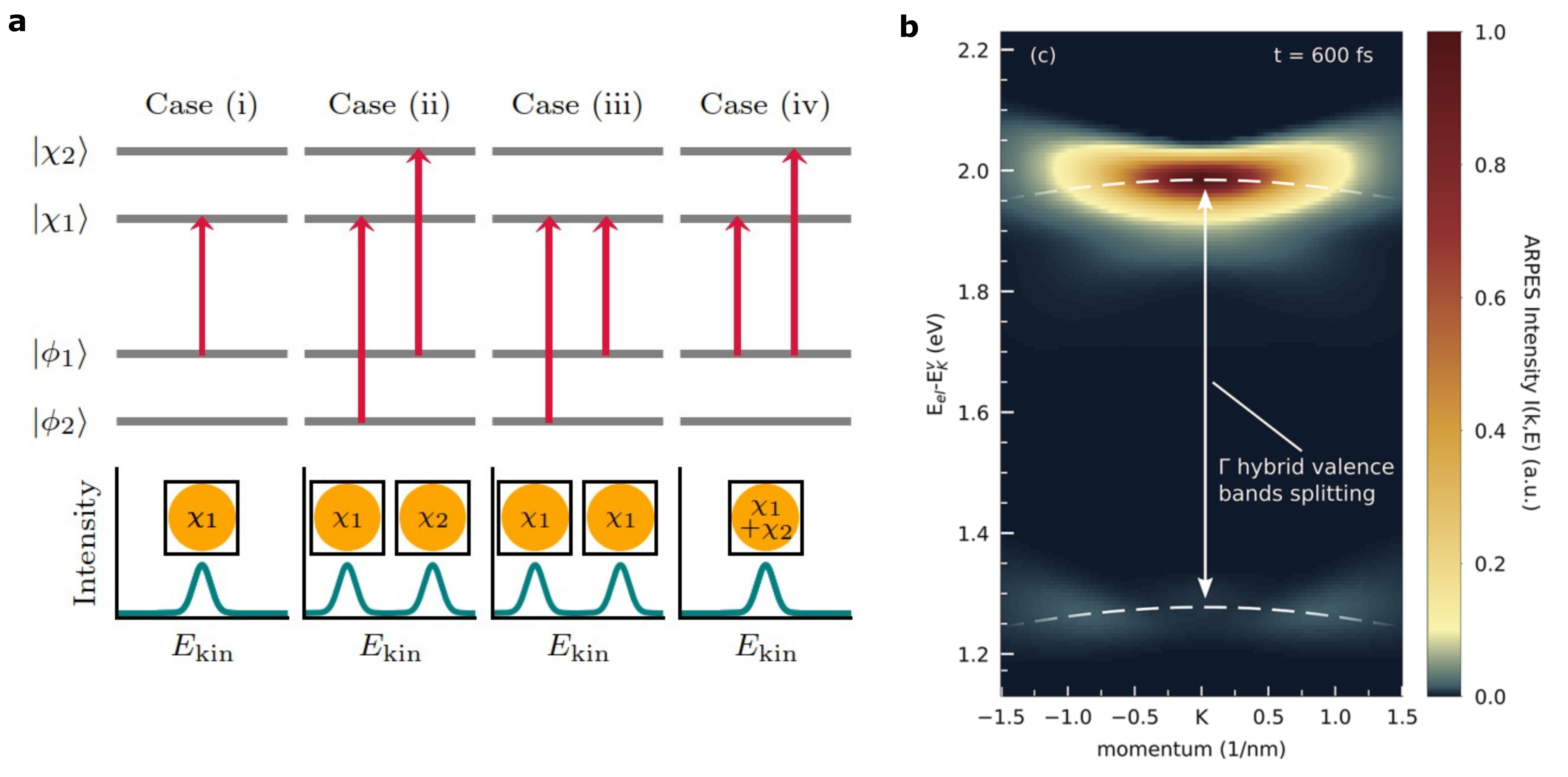}
    \caption{The multi-orbital nature of excitons can be identified in the photoemission spectroscopy experiment. 
    Kern \textit{et al.} (\textbf{a}) and Meneghini \textit{et al.} (\textbf{b}) found that multi-orbital excitons lead to more complex trARPES fingerprints, with one peak for each unique final-state binding energy of the hole. 
    Panel \textbf{a} is reproduced from ref.~\cite{Kern23prb}. Copyright 2023 by the American Physical Society. Panel \textbf{b} is reproduced with permission from ref.~\cite{Meneghini23ACSPhotonics}. Copyright 2023 American Chemical Society.}
    \label{fig:photoemission-theory-multi}
\end{figure}

Here, we first review the work of Kern \textit{et al.}~\cite{Kern23prb} and discuss the expected photoemission intensity for the case of multi-orbital excitons in an organic semiconductor. Kern \textit{et al.} realized that the complex exciton wavefunction \eqref{eq:exciton_state} can be split in four prototypical cases, where each case has a unique photoemission signature (Fig.~\ref{fig:photoemission-theory-multi}a). Case 1 is the simplest scenario, where the exciton is composed of a single valence band and a single conduction band. The excitons that have been discussed so far fit in this category. Beyond case 1, we find exciton wavefunctions that are composed of multiple elementary transitions between orbitals or bands. Case 2 specifies an exciton that is composed of two elementary transitions, namely from the second-highest occupied molecular orbital (HOMO-1) to the LUMO and from the HOMO to the second-lowest unoccupied molecular orbital (LUMO+1). This exciton can be considered as a coherent superposition of a HOMO-1 - LUMO electron-hole pair and a HOMO - LUMO+1 electron-hole pair. Case 3 and 4 are also composed of two elementary transitions, however, in these cases, the electron (case 3) or the hole (case 4) are uniquely occupying a single orbital. Of course, even more complicated wavefunctions are possible within equation~\eqref{eq:exciton_state}, but these can be expressed as linear combinations or extensions of the cases 1-4.

The next step is to calculate the characteristic photoemission signature for these individual cases. At first, it is instructive to consider energy conservation in the photoemission process, i.e., equation ~\eqref{eq:enmomconservation}, which implies that the kinetic energy of the observed photoelectron is related to the final-state energy $E_\mathrm{hole}$ of the remaining hole. For case 1 (and 4), this energy is well defined: as the exciton is derived from a single orbital, it follows that the hole remains in this state after the exciton is broken. For case 2 and 3, however, multiple valence orbitals contribute to the exciton wavefunction, and the hole may remain in either of these orbitals after photoemission. Depending on the hole final state, the photoelectron then acquires more or less kinetic energy, leading to one peak in the photoelectron spectrum for each unique hole final-state energy $E_\mathrm{hole}$ (see bottom panel of Fig.~\ref{fig:photoemission-theory-multi}a). 

Kern \textit{et al.} went beyond this first step, and calculated the complete photoelectron spectrum for weakly-interacting systems  in the plane-wave approximation, including the amplitude of the individual peaks in the photoelectron spectrum and, crucially, the momentum distribution. For an exciton with energy $E_{\rm exc}$ that is photoemitted by a photon with energy $\hbar\omega_{\rm probe}$, the angle-resolved photoelectron spectrum is proportional to
\begin{equation}
    I(E_\mathrm{kin},\ve{k})  \propto  
    \left| \ve{A} \ve{k} \right|^2
    \sum_{v} \left| \sum_{c} X^{(m)}_{v c}  \mathcal{F}\left[\chi_{c} \right] (\ve k)\right|^2 
     \times   \delta\left(\hbar\omega_{\rm probe} - E_\mathrm{kin} - E_v + E_{\rm exc} \right).
\label{eq:trPOT}
\end{equation}
Here $\ve{A}$ is the vector potential of the incident light field, $\mathcal{F}$ the Fourier transform, $\ve{k}$ the photoelectron momentum, $E_v$ the final-state binding energy of the hole in the $v$\textsuperscript{th} valence orbital, and $E_\mathrm{kin}$ the energy of the photoemitted electron. This description directly reproduces the appearance of multiple peaks in the photoelectron spectrum for excitons with multi-orbital hole contributions. 

It is interesting to observe the effect of the electron (conduction) orbitals on the momentum distribution: in case 1, the momentum distribution is directly determined by the momentum distribution $\mathcal{F}\left[\chi_{c}\right] (\ve k)$ of the electron orbital. The same happens in cases 2 and 3: here, each peak in the photoelectron spectrum is linked to electron-hole pairs derived from a single conduction orbital, and thus the momentum distribution is given by that orbital. In case 4, however, the photoelectron spectrum contains signal at the same kinetic energy coming from multiple conduction orbitals, ($\chi_1$ and $\chi_2$ in Fig.~\ref{fig:photoemission-theory-multi}a). As described by equation~\eqref{eq:trPOT}, the photoelectron momentum distribution is then given by the coherent sum of the individual electron orbitals. Thus, trARPES is sensitive to unique interferences related to the coherent sum of electron-hole pairs that builds up a single exciton, which promises a unique insight into the correlated nature of excitons.

Meneghini \textit{et al.}~\cite{Meneghini23ACSPhotonics} derive a complementary picture of the exciton photoemission spectrum and predict a very similar phenomenon for the case of bilayer TMDs. In such systems, hybridization between the top and bottom TMD layers leads to a splitting of the valence band at $\Gamma$ into two bands with bonding and anti-bonding character, respectively. These bands can host so-called hybrid excitons, in which the hole-part of the exciton wavefunction is derived from the hybridized bands at $\Gamma$. Due to the Coulomb interaction between electron and hole, the exciton's hole is in a state that is described by a superposition of band contributions from two valence bands from the neighboring layers, and after the breakup of the exciton, the exciton's hole can either remain in the band with bonding or with anti-bonding character. In this manner, very similar to case 3 in Fig.~\ref{fig:photoemission-theory-multi}a, the model predicts a two-peak photoemission structure, where the energy difference between both photoemission signatures is given by the energy splitting of the valence bands at the $\Gamma$ valley, i.e., the energy position where the single-particle holes remain the sample (Fig.~\ref{fig:photoemission-theory-multi}b).

\section{Fingerprints of excitons in time-resolved momentum microscopy}
\label{sec:excitons-in-pes}

The last few years have shown tremendous progress in the characterization of excitons in energy- and momentum-resolved photoemission spectroscopy. In this section, selected examples on how photoemission signatures of excitons could be identified in two-dimensional and organic semiconductors are provided. Thereby, we highlight how the momentum-resolved photoelectron detection scheme has facilitated the direct characterization of the energy landscape of bright and dark excitons.

\subsection{Spectroscopy of intralayer, interlayer, and hybrid excitons in TMDs}

Excitons with a large oscillator strength are routinely studied in all-optical spectroscopies. Since 2020, momentum microscopy experiments could strongly enhance the current understanding as they provided direct access to excitons with a vanishing oscillator strength. In this section, we review the recent experimental results that reported the observation of intralayer, interlayer, and hybrid excitons in exfoliated mono-~\cite{Madeo20sci, Wallauer21nanolett, Man21sciadv, kunin23prl} and twisted bilayer~\cite{Karni22nat, Schmitt22nat, Bange232DMaterials, bange24SciAdv, Schmitt23arXiv} TMDs. 

\subsubsection{Momentum-direct and -indirect excitons in monolayer TMDs}

\begin{figure}[b]  
    \centering
    \includegraphics[width=.95\linewidth]{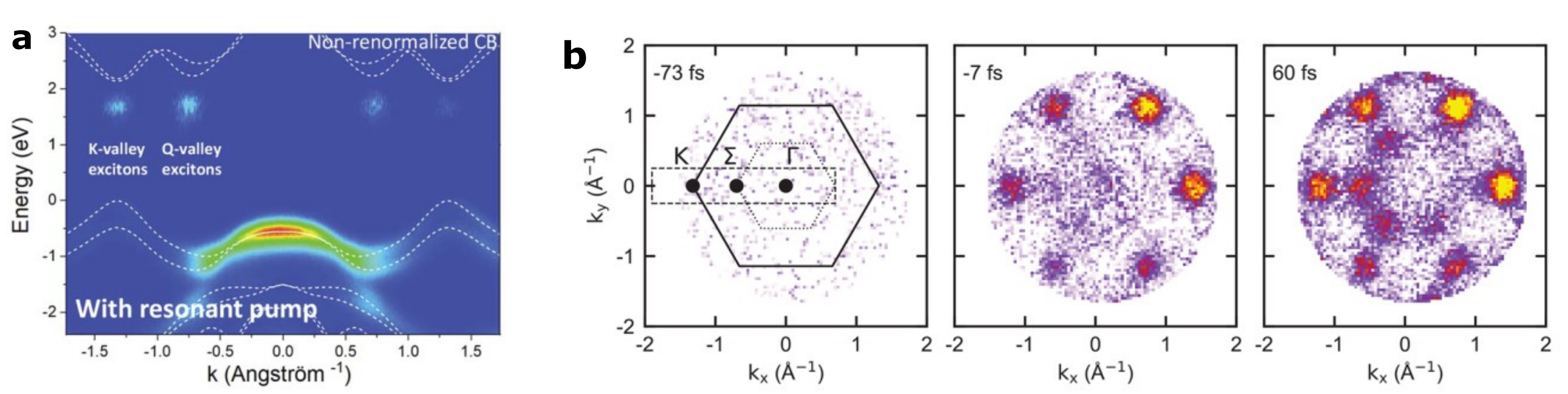}
    \caption{Energy- and momentum-resolved photoemission spectra collected on exfoliated \textbf{a} monolayer WSe$_2$ and \textbf{b} monolayer WS$_2$.
    \textbf{a} Photoemission spectral weight originating from the breakup of excitons is detected at the K and the $\Sigma$ (Q) valley within the single-particle band gap.
    \textbf{b} Pump-probe delay dependent momentum-maps indicating the optical excitation of K excitons at the K valley and the subsequent formation of dark $\Sigma$ excitons at the $\Sigma$ valley.
    Panel \textbf{a} from ref.~\cite{Madeo20sci}. Reprinted with permission from AAAS. Panel \textbf{b} reprinted with permission from ref.~\cite{Wallauer21nanolett}. Copyright 2021 American Chemical Society.
    }
    \label{fig:energymomenta-monolayer}
\end{figure}

Madéo \textit{et al.}~\cite{Madeo20sci} (Fig.~\ref{fig:energymomenta-monolayer}a) and Wallauer \textit{et al.}~\cite{Wallauer21nanolett} (Fig.~\ref{fig:energymomenta-monolayer}b) have reported the observation of photoemission spectral weight originating from excitons in momentum microscopy experiments performed on exfoliated monolayer WSe$_2$ and monolayer WS$_2$, respectively. In their work, single-particle photoelectrons being emitted from excitons have been detected at energies above the valence band maximum at the K and the $\Sigma$ valley of the hexagonal Brillouin zone, and were identified as signatures of the A-excitons and the dark $\Sigma$-excitons (Fig.~\ref{fig:energymomenta-monolayer}a,b; $\Sigma$ valley also termed Q or $\Lambda$ in literature). Fig.~\ref{fig:energymomenta-heterobilayer}c illustrates the involved excitons in an electron-hole picture for monolayer WSe$_2$: The photoemission yield at the $\Sigma$ valley originates from intralayer $\Sigma$-excitons, where the exciton's electron- and hole-components reside at the $\Sigma$ valley and the K valley, respectively. Importantly, as the exciton's components reside in different valleys of the Brillouin zone separated by momentum $\ve{w}$, those excitons cannot recombine in a radiative process and are thus termed optically dark (cf. Fig.~\ref{fig:break-exciton}d). For the case of the spectral weight at the K valley, two different excitons can contribute (Fig.~\ref{fig:energymomenta-heterobilayer}c): The optical excitation leads to the formation of excitons, where the electron- and the hole-component reside in the same K (or K$^\prime$) valley, i.e., the exciton's components have the same valley index. However, subsequent intervalley thermalization processes can lead to the formation of the energetically more favorable K-K$^\prime$ excitons, where the electron- and the hole-component have different valley indices~\cite{Selig182Dmat, Lindlau18natcom}. In consequence, these excitons are again termed optically dark due to the finite $\ve{w}$ that separates the electron- from the hole-component. It has to be noted that, as the energy difference between the K-K$^\prime$ (K$^\prime$-K) excitons and the K-K (K$^\prime$-K$^\prime$) excitons is small, it is not straightforwardly possible to spectroscopically differentiate these excitons in the photoemission experiment. If one is interested in these excitons, it is necessary to repeat the photoemission experiments in a valley-selective approach using circular polarized pump pulses~\cite{Bertoni16prl, kunin23prl}, as reported by Kunin \textit{et al.} for the case of monolayer WS$_2$~\cite{kunin23prl}.

Based on the data shown in Fig.~\ref{fig:energymomenta-monolayer}, Madéo \textit{et al.}~\cite{Madeo20sci} and Wallauer \textit{et al.}~\cite{Wallauer21nanolett} had the major task to unambiguously discriminate photoelectrons that result from the break-up of excitons and those that are emitted from the conduction band (cf. energy diagram in Fig.~\ref{fig:break-exciton}b). For this, the authors reported several arguments: First, it could be shown that photoelectrons originating from excitons are indeed detected at energies below the conduction band minimum, consistent with expectations from equation~(\ref{eq:enmomconservation}). Thereby, the energy of the conduction band was estimated by trARPES experiments where the pump photon energy was tuned to energies larger than the single-particle band gap, as has also been carried out in a report by Dong \textit{et al.}~\cite{Dong20naturalsciences} on bulk WSe$_2$. Second, excitation-energy-dependent experiments showed a clear enhancement of photoemission yield for the case that the excitation energy is tuned into resonance with the exciton energy~\cite{Madeo20sci}.

\subsubsection{Hybrid excitons in heterobilayer TMDs}

\begin{figure}[b]  
    \centering
    \includegraphics[width=\linewidth]{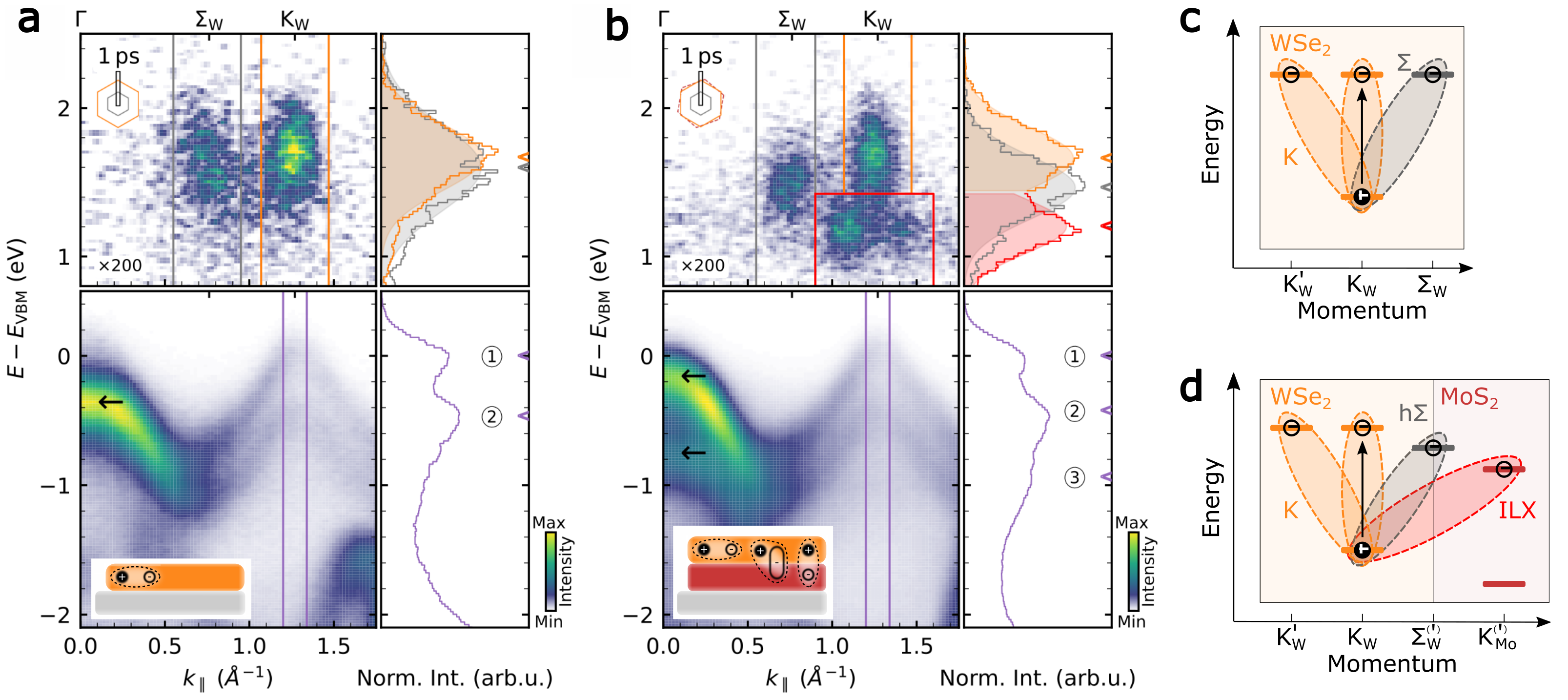}
    \caption{Energy- and momentum-resolved photoemission data collected on monolayer WSe$_2$ (\textbf{a}) and twisted heterobilayer WSe$_2$/MoS$_2$ (\textbf{b}).
    \textbf{a} In the occupied regime, the electronic band structure of monolayer WSe$_2$ is characterized by the spin-split valence bands at the K valley and the energetically lower lying valence band maxima at the $\Gamma$ valley (horizontal arrow). Photoemission signatures of optically bright and dark intralayer excitons are detected at higher photoelectron energies (cf. EDCs taken in the colored regions-of-interest).
    \textbf{b} Interlayer hybridization leads to a renormalization of the valence band structure and the formation of a split valence band at the $\Gamma$ valley (horizontal arrows). At higher energies, photoemission yield from intralayer K- (orange) and hybrid h$\Sigma$ (grey) excitons, as well as interlayer ILX (red) is detected.
    \textbf{c, d} Electron-hole picture of the low energy exciton landscape of monolayer WSe$_2$ (\textbf{c}) and heterobilayer WSe$_2$/MoS$_2$ (\textbf{d}).
    Panels adopted from from ref.~\cite{Bange232DMaterials} under Creative Commons Attribution License 4.0 (CC BY).
    }
    \label{fig:energymomenta-heterobilayer}
\end{figure}

So far, we have focused on monolayer TMDs, where the exciton's electron- and hole-components necessarily reside in the same layer. Intriguingly, the situation can become much more interesting by stacking two semiconducting TMDs on-top of each other. For example, it is possible to fabricate an atomically thin p-n junction~\cite{Hong14natnano, Lee14natnano}, i.e., a heterostructure with a type-II band alignment~\cite{Jin18natnano}. In such a heterostructure, the electronic band structure and the concomitant energy landscape of excitons depends on the composition of the heterostructure and other parameters such as the twist-angle and the emergent moiré potential. For example, interlayer hybridization between two neighboring TMDs in a bilayer, as compared to a monolayer, leads to an energetic splitting of the hybridized bands and, in consequence, to a transition from a direct to an indirect semiconductor~\cite{Mak10prl, Splendiani10nanolett}. In static ARPES experiments, the formation of hybrid electronic bands can be observed, e.g., when comparing the valence bands at the $\Gamma$ valley for the case of a mono- and bilayer TMD~\cite{Wilson17sciadv, Wencan13prl}, as visualized for the case of monolayer WSe$_2$ and heterobilayer WSe$_2$/MoS$_2$ in Fig.~\ref{fig:energymomenta-heterobilayer}a,b (black horizontal arrows).

In the case of the corresponding exciton energy landscape, interlayer hybridization will also lead to a renormalization of the exciton energies and the formation of so-called hybrid excitons, whose electron- and/or hole-components can be delocalized over both monolayers (inset Fig.~\ref{fig:energymomenta-heterobilayer}b)~\cite{Meneghini22naturalsciences, Meneghini23ACSPhotonics, brem2020hybridized, Wang17prb}. Depending on the degree of delocalization of the exciton's components across the interface, it is insightful to distinguish between three types of excitons, 
based on the trARPES data obtained from monolayer WSe$_2$ and heterobilayer WSe$_2$/MoS$_2$~\cite{Bange232DMaterials}. Figure~\ref{fig:energymomenta-heterobilayer}a and \ref{fig:energymomenta-heterobilayer}b show energy- and momentum-resolved photoemission data along the $\Gamma$-$\Sigma$-K$_{\rm W}$ crystal direction collected at a pump-probe delay of 1~ps. At energies $E - E_{\rm VBM}<0$~eV, the valence bands of WSe$_2$ and MoS$_2$ are detected. At higher energies, photoemission signatures from the (i) intralayer K-excitons (orange), (ii) interlayer excitons (red) and (iii) hybrid h$\Sigma$ excitons (gray) are found. Before describing the photoemission signatures in detail, it is important to note that the trARPES experiment performed with linear polarized pump pulses cannot differentiate between excitons where the electron- or the hole-component reside at high-symmetry points with the same or different valley indices~\cite{kunin23prl, Bertoni16prl}. 

First, the excitation of WSe$_2$/MoS$_2$ resonant to the optical band gap leads to the formation of (i) intralayer A1s-excitons that are fully localized in a single TMD layer and where the electron- and the hole-components reside in the K (or K$^\prime$) valleys. The dominant intralayer character of these excitons is due to the fact that interlayer hybridization is negligible at the K (K$^\prime$) valley of semiconducting TMDs~\cite{Wang17prb, brem2020hybridized, Wallauer20prb, Liu20prb}. In consequence, photoemission spectral weight of the intralayer excitons is found one exciton energy E$_{\rm exc}$ above the WSe$_2$ valence band maximum, where the exciton's hole remains in the sample after the break-up of the exciton. Intriguingly, a consequence of the vanishing degree of interlayer hybridization can be directly seen in experiment when comparing the photoemission spectra collected from monolayer WSe$_2$ and heterobilayer WSe$_2$/MoS$_2$: The exciton energies of the K-excitons are identical in the mono- and in the heterobilayer sample (cf. orange energy distribution curves (EDCs) in Fig.~\ref{fig:energymomenta-heterobilayer}). Second, after the excitation of the A1s-excitons, intra- and intervalley scattering events lead to the formation of (ii) interlayer excitons (ILX) via the exciton cascade K$\rightarrow$h$\Sigma\rightarrow$ILX~\cite{Schmitt22nat, Bange232DMaterials, Meneghini22naturalsciences}. For ILX, the exciton's electron- and hole-components reside in the K valleys of MoS$_2$ and WSe$_2$, respectively. As interlayer hybridization is negligible at the K valleys, the exciton's electron- and hole-components must be fully localized in one and the other single TMD layer, respectively. 
Third, in-between the excitation of (i) intralayer excitons and the build-up of (ii) ILX, (iii) hybrid excitons are formed, where the exciton's electron and/or hole components are delocalized between both layers. As interlayer hybridization goes along with a reduction of the exciton energy, the photoemission yield at the $\Sigma$ valley is found at lower photoelectron energies for the case of the WSe$_2$/MoS$_2$ heterobilayer as compared to the WSe$_2$ monolayer (grey EDCs in Fig.~\ref{fig:energymomenta-heterobilayer}). In other words, the exciton energy E$_{\rm exc}$ of the intralayer $\Sigma$ excitons in monolayer WSe$_2$ is larger than the exciton energy of hybrid h$\Sigma$ excitons in heterobilayer WSe$_2$, consistent, e.g., with photoluminescence experiments reported in refs.~\cite{Kunstmann18natphys, Merkl20natcom, Zande14nanolett, Liu14natcom}. Notably, these hybrid excitons are particularly important to mediate an efficient charge-separation process between the layers~\cite{Schmitt22nat, Bange232DMaterials, bange24SciAdv, Meneghini22naturalsciences, Meneghini23ACSPhotonics, Wang17prb, Zheng17nanolett, Wallauer20prb, Zimmermann21acsnano, Kunstmann18natphys, Policht23natcom}.


\subsection{Momentum-fingerprints of excitons in photoemission spectroscopy}

Until now, we have mainly focused on the spectral information in the photoemission experiment. The momentum-coordinate of the photoelectrons has only been of interest to differentiate between momentum-direct and -indirect excitons, i.e., to discriminate excitons with a different $\ve{w}$. Next, we review the recent efforts to evaluate the momentum-resolved photoemission intensity from a specific exciton in order to show that not only information on the exciton's electron component, but also on the exciton's hole can be extracted.


\subsubsection{Hole-like energy-momentum dispersion of the exciton's electron photoemission yield}

\begin{figure}[t]  
    \centering
    \includegraphics[width=.5\linewidth]{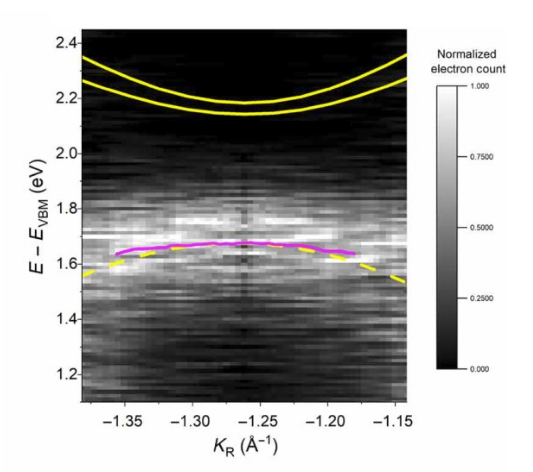}
    \caption{Hole-like energy-momentum fingerprint of photoemission spectral weight originating from the break-up of intralayer excitons in monolayer WSe$_2$. Panel reproduced from ref.~\cite{Man21sciadv} under Creative Commons Attribution License 4.0 (CC BY).}
    \label{fig:VBcurvature}
\end{figure}

The multi-dimensional data collection scheme of the momentum microscopy experiment facilitates the evaluation of the energy-momentum dispersion of the spectral weight that originates from the breakup of excitons. Figure~\ref{fig:VBcurvature} shows such data collected from monolayer WSe$_2$ by Man \textit{et al.}~\cite{Man21sciadv}. In this work, the authors have been able to identify a photoemission signature that shows a hole-like energy-momentum dispersion, as indicated by the magenta line in Fig.~\ref{fig:VBcurvature}. The data of this work was collected at a pump-probe delay of 500~fs and at liquid-nitrogen temperature, so that a sufficiently small exciton temperature was reached with excitons with a center-of-mass momentum of mainly $Q=0$. Notably, the authors found a photoemission signature with a hole-like energy-momentum-dispersion, which cannot result from photoelectrons that result from the conduction band and would thus mimic its electron-like energy-momentum dispersion~\cite{Dong20naturalsciences}. In contrast, consistent with theoretical predictions~\cite{Ohnishi18, Rustagi18prb, Christiansen19prb, Steinhoff2017natcom, Meneghini23ACSPhotonics, Kern23prb, Perfetto16prb}, Man \textit{et al.} explain the hole-like energy-momentum dispersion to result from the break-up of $Q=0$ excitons and the conservation of energy as single-particle photoelectrons are detected and single-particle holes remain in the sample.

\subsubsection{Interlayer excitons in the presence of the moiré potential}

\begin{figure}[b]  
    \centering
    \includegraphics[width=.95\linewidth]{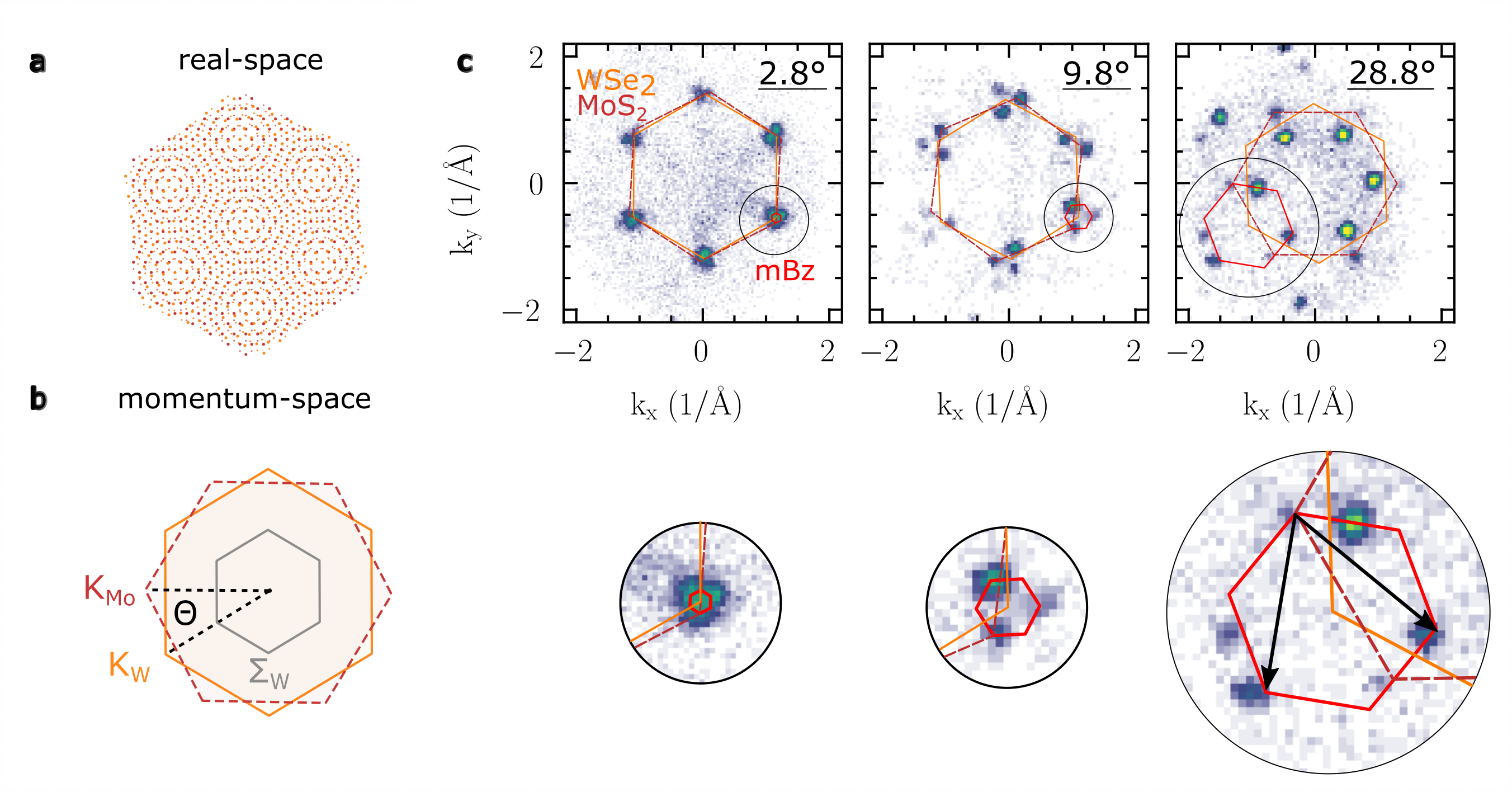}
    \caption{Moiré superlattice hallmark imprinted onto the momentum-resolved photoemission signal from interlayer excitons.
    \textbf{a} The rotational misalignment of two hexagonal lattices leads to the formation of an emergent superstructure in real-space.
    \textbf{b} Translated to momentum-space, the moiré superlattice results from the rotational misalignment of the hexagonal Brillouin zones by a twist-angle $\Theta$.
    \textbf{c} Momentum-momentum-resolved photoemission data originating from the break-up of interlayer excitons taken on  WSe$_2$/MoS$_2$ heterostructures with twist angles of 2.8°, 9.8°, and 28.8° (from left-to-right). The momentum map is taken at the energy of the interlayer exciton (cf. red EDC in Fig.~\ref{fig:energymomenta-heterobilayer}b). The top row shows the full momentum-range that is accessible in the photoemission experiment, the bottom row shows blow-ups of the black circled areas. With increasing twist-angle, the length of the moiré lattice vectors (black arrows) and the size of the moiré mini Brillouin zone (red hexagon) becomes elongated. The 9.8° and 28.8° data is reproduced and adopted from ref.~\cite{Schmitt22nat} (Copyright 2022, Nature Publishing Group) and ref.~\cite{Schmitt23arXiv}, respectively. The 2.8° data was taken in the Göttingen photoemission laboratory by D. Schmitt, J.P. Bange, and W. Bennecke from a sample prepared by A. AlMutairi and S. Hofmann and has not been published elsewhere.
    }
    \label{fig:moireexciton}
\end{figure}

The rotational misalignment of multiple two-dimensional material sheets can be used to tailor an emergent potential that manipulates the energy landscape of a TMD heterostructure in real-space, i.e., the so-called moiré superlattice. Schematically, this is visualized in Fig.~\ref{fig:moireexciton}a, where two hexagonal lattices are slightly twisted with respect to each other, leading to a well-known interference pattern. Depending on lattice constant mismatch and the twist-angle $\Theta$, the moiré wavelength can be as large as 20~nm and the potential energy well as deep as 100~meV~\cite{Zhang17sciadv,Wu17prl}. This additional modulation of the energy landscape in real-space enables further tailoring of the material's properties and can be used, for example, to create and tune correlated electronic and excitonic phases~\cite{Wu18prl, Cao18nat, Wang20natmat, Regan20nat}. For the sake of this review, the most interesting point is that excitons can interact with the additional periodic potential energy landscape. Specifically, depending on the Bohr radius and the moiré wavelength, excitons can either be localized (confined) in a single moiré potential well or they can be delocalized over multiple moiré unit cells. Exciton confinement has been experimentally demonstrated in optical spectroscopies where a fine structure in the fluorescence spectra could be attributed to the formation of so-called moiré minibands~\cite{Alexeev19nat, Tran19nat, Seyler19nat}. Naturally, the renormalization of the energy landscape of excitons due to the formation of moiré minibands is then a function of the twist-angle $\Theta$~\cite{Alexeev19nat, Brem20nanolett}, and, hence, it is highly desirable to extract such information in a momentum-resolved photoemission experiment. In the following, we aim to review how a hallmark of the moiré potential is imprinted onto the momentum-resolved interlayer exciton photoemission signal.

To set the stage, Fig.~\ref{fig:moireexciton}b shows the consequence of the real-space moiré periodicity in momentum-space. The real-space misalignment of the monolayers translates to a rotational misalignment of the hexagonal Brillouin zones of both layers by the twist-angle $\Theta$. Naturally, this misalignment of the hexagonal Brillouin zones can then be visualized in static ARPES experiments by the misalignment of the valence bands and, potentially, a renormalization of the electronic band structure~\cite{Stansbury21sciadv, Ulstrup20sciadv, Wilson17sciadv, Gatti23prl}. Recently, we and others extended this discussion to the excitonic response of a 9.8° twisted WSe$_2$/MoS$_2$ heterostructure~\cite{Schmitt22nat}. While the energy-momentum-resolved data obtained from this sample is already described above in Fig.~\ref{fig:energymomenta-heterobilayer}b, Fig.~\ref{fig:moireexciton}c shows the corresponding momentum-momentum-resolved visualization of the photoemission data (data taken at the energy of the photoelectron energy of the interlayer excitons (ILX)). Before discussing the data in detail, one has to recapitulate what might be expected for the photoemission signal originating from the ILX residing in the twisted heterostructure. The ILX's electron- and hole-components reside in the K$_{\rm Mo}$ and K$_{\rm W}$ valley of MoS$_2$ and WSe$_2$, respectively. In the description of the energy-momentum-resolved photoemission data from excitons shown in Fig.~\ref{fig:energymomenta-monolayer} and Fig.~\ref{fig:energymomenta-heterobilayer}, a naive rule of thumb was that the single-particle photoelectrons are detected at the in-plane momenta where the exciton's electron component resides in the electron-hole picture (cf. Fig.~\ref{fig:break-exciton}d), as first shown by Madéo \textit{et al.} and Wallauer \textit{et al.} for intralayer excitons in monolayer TMDs~\cite{Wallauer21nanolett, Madeo20sci}. In consequence, for the ILX, one would expect to observe photoemission yield at the K$_{\rm Mo}$ (and K$_{\rm Mo}^\prime$) valleys. However, the momentum-resolved photoemission data shown in Fig.~\ref{fig:moireexciton}c is more complex and shows a three-fold structure. To understand this ILX momentum fingerprint, the moiré mini Brillouin zone (mBz) that is built-up by the rotationally misaligned reciprocal lattice vectors of WSe$_2$ and MoS$_2$ (red hexagons in Fig.~\ref{fig:moireexciton}c) in combination with an extended zone scheme (see refs.~\cite{Ahn18sci, Koshino15njp}) needs to be considered~\cite{Schmitt22nat}. In such an analysis, it is found that photoemission yield of the ILX is not only detected at the momentum of the exciton's electron location (i.e., the K$_{\rm Mo}^{(\prime)}$ valley), but also at those momenta accessible via Umklapp processes with reciprocal lattice vectors of the layer where the exciton's hole resides. Hence, the photoemission momentum-structure of the ILX shows a hallmark of the moiré superlattice, and it contains direct information on the exciton's electron- and hole-components. In reciprocal space, an increasing twist-angle $\Theta$ elongates the moiré lattice vectors (black arrows in Fig.~\ref{fig:moireexciton}c), leading to an increasing size of the mBz (red hexagon). Naturally, by Fourier analogy, this goes along with a decreasing size of the real-space moiré wavelength. 

From the twist-angle-dependent data in Fig.~\ref{fig:moireexciton}c, two conclusions can be drawn: First, trARPES is capable to experimentally access ILXs in a nearly 30° misaligned heterostructure, i.e., from a sample where the momentum offset $\ve{w}$ of the exciton's electron- and hole-component is maximal and only leads to vanishing fluorescence signal~\cite{Nayak17acsnano}. Even more important, second, the momentum-resolved photoelectron detection scheme can be used to directly visualize the transition from a twist-angle regime where the formation of moiré minibands can be neglected (large $\Theta$, i.e., 9.8° and 28.8° in Fig.~\ref{fig:moireexciton}c) to the regime where the twist-angle is sufficiently small such that Umklapp processes with the reciprocal moiré superlattice vectors leads to overlap between the excitonic bands and thus the formation of moiré mini bands (small $\Theta$, i.e., 2.8° in Fig.~\ref{fig:moireexciton}c). For such small twist-angles $\Theta$, one gets into a regime where excitons can be confined in a single moiré potential well~\cite{Alexeev19nat, Brem20nanolett}. In this manner, it is important to highlight the work of Karni \textit{et al.}~\cite{Karni22nat}, who reported momentum microscopy experiments on a 2.2° twisted WSe$_2$/MoS$_2$ heterostructure and could reconstruct the full exciton wavefunction that is confined in a single moiré potential well (see subsection~\ref{subsection:realspace_reconstruction}).


\subsection{Photoelectron spectrum of multi-orbital excitons}
\label{subsec:multiorbital_experiment}

\begin{figure}[b] 
    \centering
    \includegraphics[width=\linewidth]{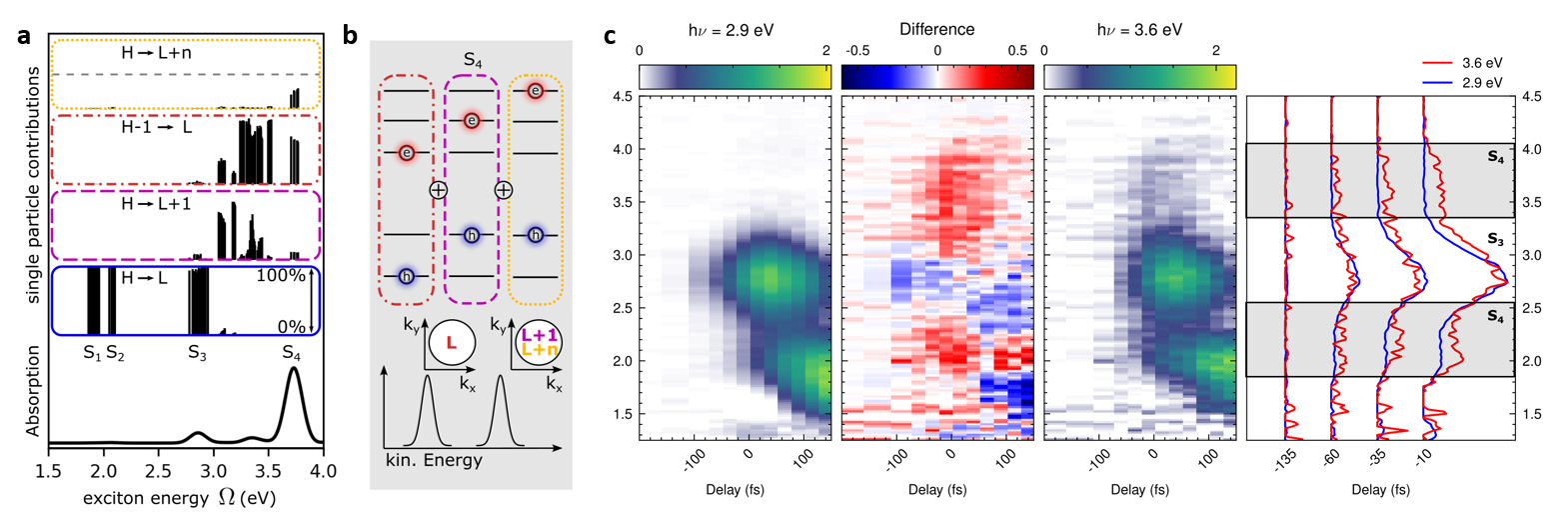}
    \caption{Multi-orbital excitons in the organic semiconductor C$_{60}$. 
    \textbf{a} A calculation in the combined many-body interaction frameworks of the $GW$ approach and the Bethe-Salpeter equation predicts that multi-orbital excitons are prevalent in C$_{60}$. 
    \textbf{a} In particular, an exciton band around 3.6~eV excitation energy (`S$_4$') is identified, in which both the HOMO and the HOMO-1 contribute significantly to the hole character of the exciton (case 2 in Fig.~\ref{fig:photoemission-theory-multi}a). This leads to the prediction of two photoemission features at 3.6~eV above the HOMO and 3.6~eV above the HOMO-1 (= 2.2~eV above the HOMO), respectively. 
    \textbf{c} A challenge in the measurement of this double signature, however, is the short lifetime of the S$_4$ exciton band. By performing momentum microscopy measurements on the sub-100-fs timescale at two distinct pump energies, it is possible to identify an enhancement of the photoemission yield that is only present during excitation of the S$_4$ exciton band (gray shading in \textbf{c}, right panel). Figure adopted from ref.\cite{Bennecke23arxiv} under Creative Commons Attribution License 4.0 (CC BY).}
    \label{fig:multiorbitalC60}
\end{figure}

In the last part of this section, we review the recent experimental progress concerning the probing of multi-orbital excitons in momentum microscopy. As discussed in subsection \ref{subsec:multiorbital_theory} and shown in Fig.~\ref{fig:photoemission-theory-multi}, such excitons are predicted to yield multi-peaked signals in trARPES, appearing at kinetic energies separated by the difference in binding energy of the different hole orbitals that contribute to the exciton. 

Recently, we have collaborated in a time-resolved momentum microscopy study of exciton dynamics in multilayer C$_{60}$~\cite{Bennecke23arxiv}, where the experimental data is compared to theoretical exciton spectra calculated within the many-body interaction frameworks of the $GW$ approach and the Bethe-Salpeter equation. Specifically, it is predicted that the lower-energy excitons, which have been studied in trARPES before~\cite{Stadtmuller:2019fe, emmerich_ultrafast_2020}, are derived from the (multiply degenerate) HOMO and LUMO levels, while higher-lying excitons with an exciton energy of 3~eV and more are of increasingly multi-orbital nature (Fig.~\ref{fig:multiorbitalC60}a). For example, at an exciton energy of 3.6~eV, a band of excitons is identified which shows clear multi-orbital nature where the exciton's hole-component is spread over the HOMO and HOMO-1 orbitals (S$_4$-exciton). In the pump-probe delay evolution of momentum-integrated energy distribution curves collected in the trARPES experiment, the double-peak structure could indeed be identified (Fig.~\ref{fig:multiorbitalC60}c, grey shading), as predicted for an exciton of type (ii) as shown in Fig.~\ref{fig:photoemission-theory-multi}a and for the case of the C$_{60}$ molecule in Fig.~\ref{fig:multiorbitalC60}b. In this manner, these results provide an experimental confirmation of the predictions in refs.~\cite{Kern23prb} and \cite{Meneghini23ACSPhotonics}, and, more generally, showcase how trARPES can experimentally disentangle the multi-orbital contributions to a single exciton band.

At this point, it is important to note that it is perhaps a bit surprising that such multi-orbital excitons were not studied in detail before. The general theoretical framework of excitons allows such excitations \cite{krylov_orbitals_2020, martin_natural_2003, hammon_pump-probe_2021}, and following the recent results \cite{Meneghini23ACSPhotonics, Kern23prb, Bennecke23arxiv, Neef23nat} it is clear that such excitons are expected to arise in TMDs as well as organic semiconductors. Although research on this topic is still in a very early stage, it is possible to identify two possible reasons why such multi-orbital excitons have not been observed before: First, the work on organic semiconductors suggests that multi-orbital excitons are more likely to occur at higher photon energies above 3~eV~\cite{Kern23prb, Bennecke23arxiv}. At these energies, absorption of two photons from the pump pulse commonly suffices to photoemit an electron above the vacuum level, which renders two-color trARPES studies rather challenging. In fact, precisely this process limited the applicable pump fluence in the study on C$_60$ discussed above~\cite{Bennecke23arxiv}. On the other hand, the work by Meneghini \textit{et al.}~\cite{Meneghini23ACSPhotonics} illustrates another experimental challenge in the detection of multi-orbital excitons: the relative strength of the photoemission features is determined by the relative occupation of the different hole states, meaning that in many cases, one feature will be significantly brighter than the other (compare intensity difference of the higher- and lower-energy-peak in Fig.~\ref{fig:photoemission-theory-multi}b). The observation of multi-orbital excitons therefore requires a good signal-to-noise ratio in trARPES, especially if the splitting of the hole states is smaller than or comparable to the energy resolution of the experiment.



\section{Photoemission orbital tomography of excitons}
\label{sec:progressOS}

In this section, we switch the analysis of exciton photoemission from momentum-space to real-space, and discuss how the real-space exciton wavefunction can be reconstructed with \AA{}ngström-level resolution from momentum-resolved photoemission data. This is possible by means of photoemission orbital tomography (POT), a technique that was first developed for the imaging of molecular orbitals by static ($k_x, k_y$)-resolved ARPES~\cite{Puschnig09sci}. POT is based upon the plane-wave model of photoemission, which states that the final-state of the photo-excited electrons can be well approximated by a plane wave. For two-dimensional systems such as TMDs and also flat organic semiconductors such as pentacene, this leads to the conclusion that the two-dimensional photoelectron momentum distribution I($k_x$,$k_y$) is proportional to the Fourier transform of the single-particle real-space molecular orbital $\mathcal{F} [ \psi\left(r_x,r_y \right)]$ up to a smoothly-varying factor $\ve{A}\ve{k}$, i.e.,
\begin{equation}
    I(k_x,k_y)\propto \left|\mathcal{F} [ \psi\left(r_x,r_y \right) ]\right|^2.
    \label{eq:FTrelationship}
\end{equation}

While the accuracy of the plane wave model has been topic of discussion~\cite{bradshaw_molecular_2015, egger_can_2019, kern_simple_2023, dauth_orbital_2011}, it has been very successfully applied in the last years. For instance, the plane-wave model enables to calculate accurate predictions of the photoemission momentum distribution of occupied molecular orbitals that have been calculated by density functional theory. In POT studies, the comparison of these predictions to experimental data allow an exceptional insight into the (electronic) structure of well-ordered molecular thin films and metal-organic hybrid interfaces~\cite{yang_momentum-selective_2022, kliuiev_combined_2019}. Perhaps the most intriguing application of the plane-wave model is, however, to image the molecular orbital from experimental ARPES data alone. Similar to X-ray coherent diffractive imaging~\cite{miao_extending_1999}, numerical phase retrieval algorithms can be used to determine the phase of the photoelectron distribution, and thereby enable to calculate an image of the molecular orbital by an inverse Fourier transform~\cite{Puschnig09sci, luftner_imaging_2014, weis_exploring_2015, jansen_efficient_2020}. 


\subsection{Time-resolved photoemission orbital tomography}

Although photoemission orbital tomography was first developed as a static technique, the potential to employ POT in a time-resolved spectroscopy has been realized early on, and it has been a long-standing goal of the scientific community. Similar to the measurement of exciton dynamics in exfoliated 2D materials, it was the development of momentum microscopy in combination with high-repetition rate EUV beamlines that enabled time-resolved photoemission orbital tomography (trPOT) to be realized. Time-resolved momentum microscopy for POT was first harnessed by Wallauer \textit{et al.}~\cite{Wallauer20sci}, who realized that trPOT not only requires an efficient measurement scheme, but also an excited state that lives long enough to provide sufficient signal on top of the strong photoemission signal from occupied molecular orbitals. Wallauer \textit{et al.} created such a long-lived state by decoupling the organic adsorbate layer (here perylene-3,4,9,10-tetracarboxylic dianhydride, PTCDA) from the metallic Cu(001) surface using a thin oxide layer. 

As in the case of any seminal work in a new field of research, the work by Wallauer \textit{et al.} also raises some important questions. For example, when applied to excitonic states, does trPOT have the potential to shed light on the dynamics and structure of the Coulomb-correlated electron-hole pairs? Given the two-particle nature of the excitons, can trPOT access the real-space properties of both the electron and the hole? Since ground-state density functional theory does not give access to the exciton spectrum, what level of theoretical description is necessary to access real-space properties in trPOT? In subsection~\ref{subsec:multiorbital_theory}, we have already seen that new theoretical frameworks are currently being developed that connect the real-space exciton wavefunction to the momentum-resolved photoemission experiment. In the following sections, we will review how this framework has recently been exploited to access the real-space properties of excitons in TMDs and organic semiconductors alike. Before going into detail, it has to be noted that the reviewed theoretical framework (Eq.~\eqref{eq:trPOT}) is only valid for excitons consisting of a single electron and hole. The situation is much more complicated when composite excitons such as biexcitons or trions are considered~\cite{pei_many-body_2019, barbone_charge-tuneable_2018, vektaris_new_1994}, as in this case the left-behind state after photoemission can be excitonic in itself~\cite{Neef23nat}.


\subsection{Two-dimensional semiconductors: Reconstruction of the real-space wavefunction of excitons}
\label{subsection:realspace_reconstruction}

In section~\ref{sec:excitons-in-pes}, we have reviewed the energy- and momentum-resolved photoemission spectra that are collected after the the break-up of excitons. In these, the in-plane momentum-resolved photoemission intensity contains information on the electron-contribution to the exciton wavefunction, and, moreover, indirect information on the hole-component can be extracted from the momentum-fingerprint. In this section, we review how recent publications used such data and the framework of orbital tomography to reconstruct the real-space distribution of the electron-contribution to the exciton wavefunction. Since 2021, this concept has been applied to intralayer excitons in monolayer WSe$_2$~\cite{Man21sciadv}, intralayer excitons in bulk WSe$_2$~\cite{Dong20naturalsciences}, and intra- and interlayer excitons in twisted heterobilayer WSe$_2$/MoS$_2$~\cite{Schmitt22nat, Karni22nat}.

\begin{figure}[t]  
    \centering
    \includegraphics[width=.95\linewidth]{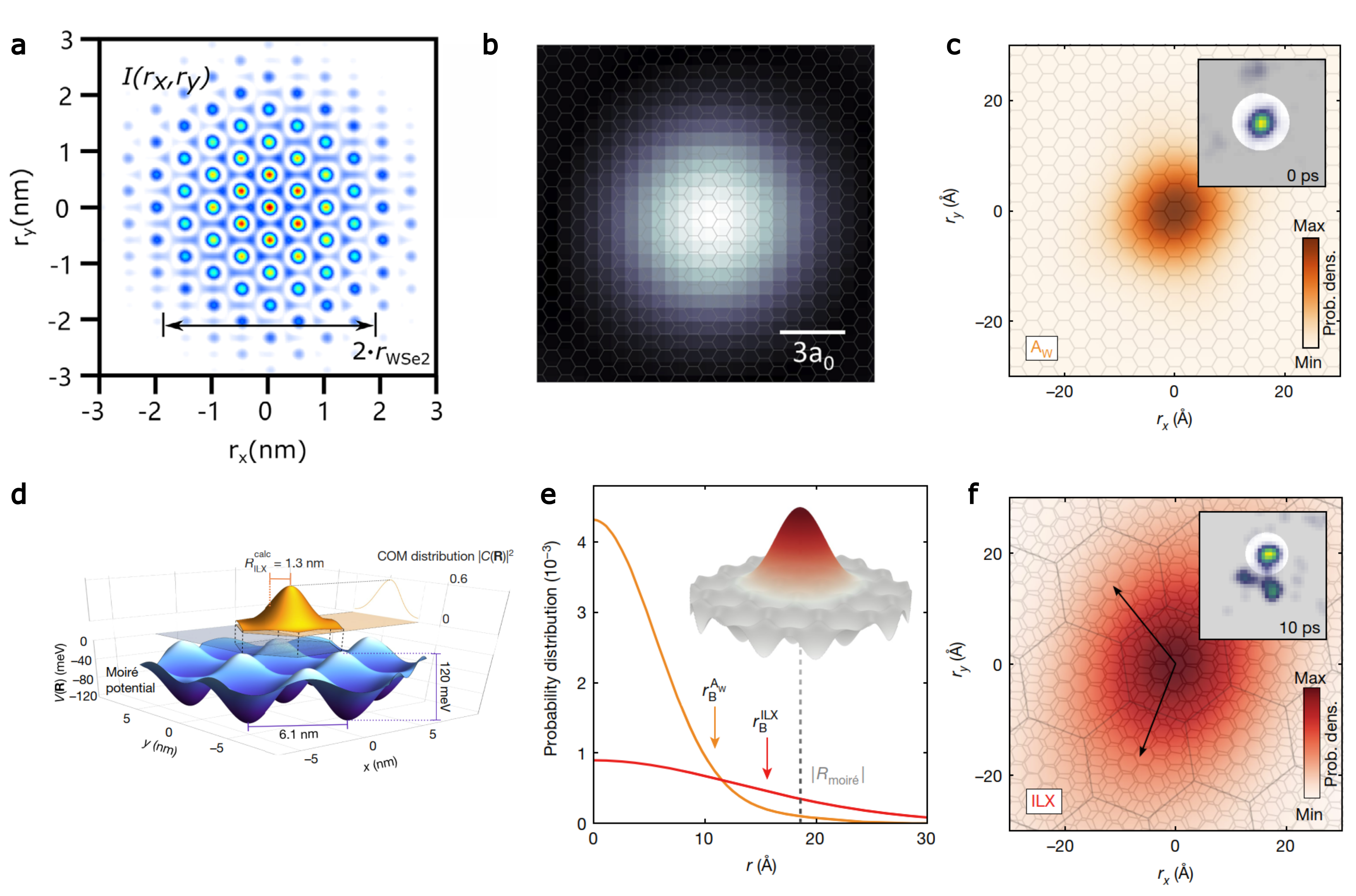}
    \caption{Reconstruction of the real-space resolved exciton wavefunction from momentum-resolved photoemission data.
    \textbf{a-c} Application of the framework of orbital tomography to reconstruct the electron-contribution to the exciton wavefunction of intralayer excitons in bulk WSe$_2$ (\textbf{a}), monolayer WSe$_2$ (\textbf{b}), and heterobilayer WSe$_2$/MoS$_2$ (\textbf{c}). The reported Bohr radii are 1.72~nm, 1.4~nm, and $1.1\pm 0.1$~nm, respectively.
    \textbf{d} Calculated real-space envelope of the interlayer exciton wavefunction in the center-of-mass coordinate on 2.2° twisted WSe$_2$/MoS$_2$.
    \textbf{e} Analysis of the exciton Bohr radius of intralayer and interlayer excitons in WSe$_2$/MoS$_2$.
    \textbf{g} Reconstruction of the real-space distribution of the electron-contribution to the exciton wavefunction of interlayer excitons in 9.8° twisted WSe$_2$/MoS$_2$.
    Panel \textbf{a} is adopted from ref.~\cite{Dong20naturalsciences} under Creative Commons Attribution License 4.0 (CC BY). Panel \textbf{b} is adopted from ref.~\cite{Man21sciadv} under Creative Commons Attribution License 4.0 (CC BY). Panels~\textbf{c,e,f} are adopted from ref.~\cite{Schmitt22nat} (Copyright by Springer Nature). Panel \textbf{d} is adopted from ref.~\cite{Karni22nat} (Copyright by Springer Nature).}
    \label{fig:realspaceTMD}
\end{figure}

Figs.~\ref{fig:realspaceTMD}a-c show the real-space probability density to find the electron-contribution to intralayer excitons if the hole-component is fixed at the center of the coordinate system. Such data is obtained by evaluating the Fourier transform (Eq~\eqref{eq:FTrelationship}) that connects the real-space wavefunction $\Psi\left(r_x,r_y \right)$ (main panels Figs.~\ref{fig:realspaceTMD}a-c) with the momentum-resolved photoemission intensity $I(k_x,k_y)$ (inset Fig.~\ref{fig:realspaceTMD}c) within the plane wave approximation and the assumption of a constant phase profile~\cite{Puschnig09sci, Schmitt22nat, Man21sciadv, Dong23natcom}. With this data at hand, it has then been possible to evaluate the exciton Bohr radius that was found to be in the order of 1-2~nm, and, therefore, extends over multiple lattice sites of the TMD (see hexagonal overlays in Fig.~\ref{fig:realspaceTMD}a-c).

Having reconstructed the electron-component to the wavefunction of intralayer excitons, in the next step, Schmitt \textit{et al.}~\cite{Schmitt22nat} and Karni \textit{et al.}~\cite{Karni22nat} focused on interlayer excitons in WSe$_2$/MoS$_2$ and the question in how far the exciton wavefunction is delocalized over multiple moiré unit cells or confined in a single moiré potential well. Schmitt \textit{et al.}~\cite{Schmitt22nat} focused on a 9.8° twisted heterostructure and found that the exciton Bohr radius is indeed - as expected from photoluminescence experiments and theory~\cite{Alexeev19nat, Brem20nanolett} - larger than the size of a moiré unit cell (Fig.~\ref{fig:realspaceTMD}e). Karni \textit{et al.}~\cite{Karni22nat} performed a similar experiment on a sample with a twist angle of 2.2° and indeed found that the exciton wavefunction is confined in a single moiré potential well (Fig.~\ref{fig:realspaceTMD}d). Notably, in the work of Karni \textit{et al.}, the authors found a very distinctive gap in the ARPES signature of the valence band that they attributed to the hole-component of the interlayer exciton. In this way, they succeeded in reconstructing both the electron- and the hole-coordinate of the exciton wavefunction (cf. equation~(\ref{eq:exciton_state})).

\subsection{Organic semiconductors: Probing the localization and delocalization of excitons in real-space} 
\label{subsec:realspace_localization}

Finally, we review the analysis of real-space exciton wavefunctions in organic semiconductors. Here, a crucial difference with the TMD semiconductors should be noted: whereas TMDs are periodic systems in which the excitons derive their wavefunction from the Bloch states of the valence and conduction bands, the electrons in organic semiconductors are more localized. Consequently, excitons in TMDs are Wannier excitons \cite{wannier_structure_1937}, while excitons in organic semiconductors are conventionally classified into Frenkel and charge-transfer (CT) excitons \cite{cudazzo_frenkel_2013, knoester_frenkel_2003, agranovich2009excitations}. Frenkel excitons have their electron and hole components on the same molecule, while the electron and hole components in a CT exciton are separated over multiple molecules. 

\begin{figure}[b]  
    \centering
    \includegraphics[width=\linewidth]{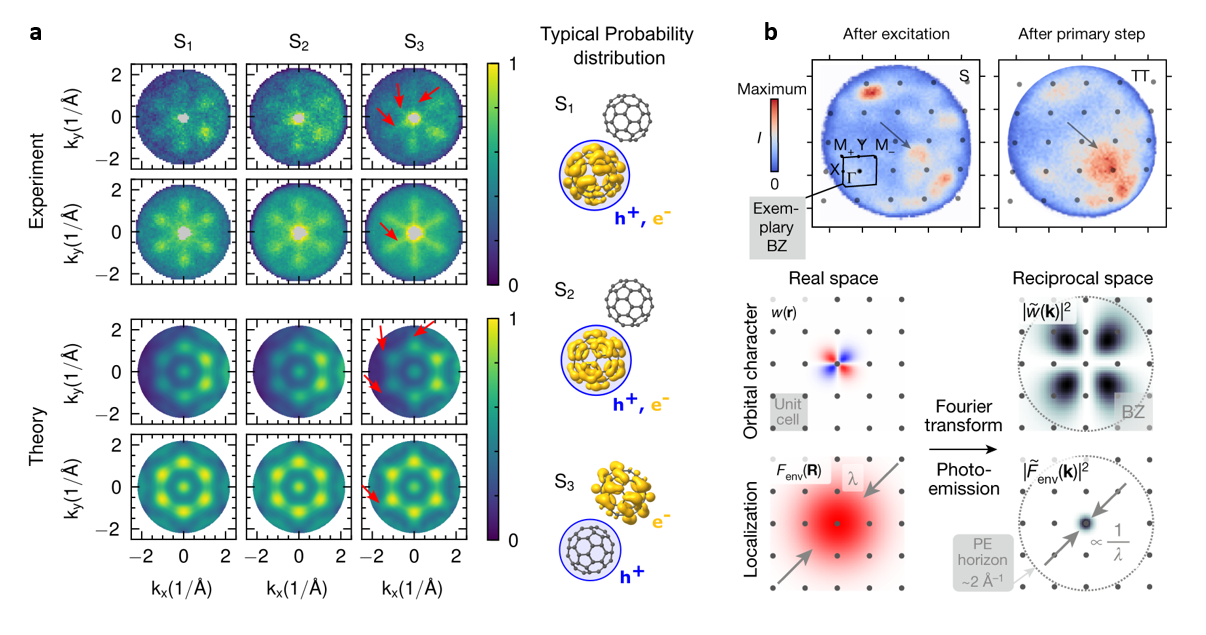}
    \caption{
    Real-space analysis of the exciton wavefunction in organic semiconductors. 
    \textbf{a} In ref.~\cite{Bennecke23arxiv}, the authors directly compare the momentum microscopy data of a C$_{60}$ multilayer to ab-initio calculations based on the $GW$+BSE framework and equation~\eqref{eq:trPOT}. Whereas the momentum fingerprints for the lower-energy S\textsubscript{1} and S\textsubscript{2} exciton bands are accurately predicted, the theory fails to fully explain the differences in the S\textsubscript{3} momentum fingerprints. These limitations are attributed to the limited size of the unit cell in the calculations. Nevertheless, the comparison of experiment and theory supports a CT-nature for the S\textsubscript{3} exciton band. 
    \textbf{b} In pentacene crystals, the optically bright singlet exciton at 1.81~eV decays by singlet fission into a pair of triplet excitons. Neef \textit{et al.}~\cite{Neef23nat} exploit the Fourier relationship between orbital and trARPES data to probe the exciton localization. They observe that the optically-excited singlet (S, top-left) has well-defined peaks in momentum, while the bitriplet (TT, top-right) is spread out; so the excited state transitions from a delocalized to a localized state. 
    Panel \textbf{a} is adapted from ref.~\cite{Bennecke23arxiv} and panel \textbf{b} is adapted from ref.~\cite{Neef23nat} under Creative Commons Attribution License 4.0 (CC BY).}
    \label{fig:real_space_trPOT_OS}
\end{figure}

The potential of momentum microscopy to shine light on the charge-transfer or Frenkel nature of an exciton is well illustrated by the recent research on crystalline C$_{60}$~\cite{Bennecke23arxiv}. Whereas CT excitons are commonly studied in multi-component systems such as hybrid organic/inorganic interfaces \cite{zhu_charge-transfer_2009, Tanda23pssa, Gonzalez22prm}, an indication for a CT exciton was recently found in a momentum-integrated time-resolved photoemission spectroscopy experiment on C$_{60}$\cite{wang_aggregates_1993, causa_femtosecond_2018, Stadtmuller:2019fe, emmerich_ultrafast_2020}. However, as this initial work on CT excitons in C$_{60}$ was performed using a hemispherical analyzer and no full momentum distributions were available at EUV energies, it was not yet possible to apply the framework of POT to access the real-space wavefunction of the excitons. This task could only be later realized in a study based on momentum microscopy experiments~\cite{Bennecke23arxiv}. Indeed, the ab-initio many-body interaction $GW$+BSE calculations in this work, plugged into equation~\eqref{eq:trPOT} and combined with the full EUV momentum microscopy data, provide the opportunity to analyze the real-space wavefunction of these excitons. Such a comparison, and exemplary visualizations of the electron-hole wavefunction, are shown in Fig.~\ref{fig:real_space_trPOT_OS}a. While this comparison was able to confirm the CT nature of the 2.8~eV exciton in multilayer C$_{60}$, the results are still limited by the computational expense of a full $GW$ and BSE calculation, which in this case was performed for a unit cell of two C$_{60}$ molecules. It is to be expected that the field of exciton POT will benefit significantly from future improvements in the computational power and methods. 

TrPOT can also provide highly valuable insight even in absence of a full theoretical description of the exciton wavefunction, as was shown in a recent study by Neef \textit{et al.}~\cite{Neef23nat}. Turning their attention to exciton dynamics in pentacene crystals, Neef \textit{et al.} applied trPOT to study singlet-fission in this material~\cite{Chan11sci, Smith10cr}. Here, the momentum-resolved detection scheme provided two crucial benefits: first, the different excitonic states, particularly the optically excited singlet state and the intermediate bitriplet exciton, each have their own unique momentum finger print. By establishing these fingerprints, Neef \textit{et al.} are able to decompose the measurement signal, and thus measure state-resolved dynamics where a decomposition purely based upon the photoelectron kinetic energy would not be possible. Second, the authors also make use of the Fourier relationship between measurement and wavefunction that is inherent to (tr)POT: as shown in Fig.~\ref{fig:real_space_trPOT_OS}b, the observed momentum fingerprints show distinct differences, with the optically excited singlet exciton displaying more pronounced peaks and the bitriplet exciton showing a more slowly-varying amplitude. From this data, Neef \textit{et al.} elucidate that the (electron-contribution to the) singlet exciton in pentacene is delocalized, while the (electron-contributions to the) bitriplet are localized to a single molecule. These results show that the framework of trPOT can provide detailed spatial information on the exciton wavefunction, even for complicated cases such as the bitriplet biexciton, where no general theoretical framework yet exists to predict the momentum-resolved photoemission signature.



\section{Summary}
\label{sec:outlook}

We have reviewed the recent efforts to make use of state-of-the-art time-resolved momentum microscopy to study excitons in two-dimensional and organic semiconductors in real- and momentum-space. By evaluating the kinetic energy and the momentum of single-particle photoelectrons originating from excitons, it is possible to quantify the exciton energy E$_{\rm exc}$, the exciton binding energy E$_{\rm bin}$, and also information on the momentum-coordinate of the exciton wavefunction. On the one side, this facilitates direct access to the long-sought after dark exciton energy landscape in two-dimensional TMDs. On the other side, this approach enables the measurement of very distinct momentum-fingerprints of excitons that could be related to the emergent moiré potential. Moreover, building on the momentum-resolved detection scheme, it has become possible to reconstruct the real-space exciton wavefunction within the framework of orbital tomography, providing thus access to the real-space wavefunction of an excited state.

\clearpage


\section{ACKNOWLEDGMENTS}
At this point, we would like to thank all colleagues that contributed to the research reviewed in the present article. First, we would like to highlight the members of the Göttingen photoemission laboratory (in alphabetical order): Jan Philipp Bange, Wiebke Bennecke, Marius Keunecke, Mattis Langendorf, Marco Merboldt, Christina Möller, Hannah Strauch, Paul Werner, David Schmitt, Daniel Steil, Sabine Steil, and Bent van Wingerden. Second, we would like to thank Ermin Malic (University of Marburg), his team members Samuel Brem and Giuseppe Meneghini, as well as Peter Puschnig (University of Graz) and his team members Christian S. Kern and Andreas Windischbacher for the fruitful experiment-theory collaborations. And third, we would like to thank Stephan Hofmann (University of Cambridge) and his team members AbdulAziz AlMutairi and Irina Chirca, Thomas Weitz (University of Göttingen) and his team members Jonas F. Pöhls and Anna Seiler, and Martin Aeschlimann, Benjamin Stadtmüller and their team member Ralf Hemm for the fruitful collaboration on sample fabrication and characterization. Finally, we thank the Deutsche Forschungsgemeinschaft (DFG, German Research Foundation) who funded this research via 432680300/SFB 1456 (project B01) and 217133147/SFB 1073 (projects B07 and B10).

\section{DISCLOSURE STATEMENT}

No potential conflict of interest was reported by the author(s).

\bibliography{bibtexfile}

\end{document}